\documentstyle[12pt,thmsa]{report}

\input tcilatex

\begin{document}

\title{Studying Bounds on Lepton Flavor Violating (LFV) Decay Processes\\
\vspace{2in} \pagenumbering{roman} \thispagestyle{empty}}
\author{Muhammad Jamil Aslam\vspace{1.0in} \\
Department of Physics \\
and \\
National Center for Physics\\
Quaid-e-Azam University\\
Islamabad 45320\\
Pakistan ~}
\date{December 2001}
\maketitle
\tableofcontents
\listoffigures

\begin{abstract}
This dissertation reviews the Standard Model formalism as well as the Lepton
Flavour Violating (LFV)\ decay processes which cause its extension, known as
the physics beyond the SM. Firstly, using the experimental bounds on three
body LFV decays, the corrosponding bounds on two body LFV decays are
reviewed. The dynamical suppression of three body LFV decays due to momentum
dependent couplings is also reviewed. Secondly, the role of the LFV decays
to explain the LSND excess is discussed in detail, for which the
experimental bounds on three body LFV\ decays, i.e. $\mu \rightarrow 3e$ are
used to constraint the coupling $\tilde{g}_{Z_{\mu e}}$, which is needed to
calculate the anamolous muon decay $\mu \rightarrow e\nu _l\bar{\nu}_l$ .
Then comparing the effective coupling of anamolous muon decay to $%
r>1.6\times 10^{-3}$\cite{9809524}, it is proved that LFV is not the correct
hypothesis to explain the LSND excess. Finally, LFV decays at loop order are
studied in Seesaw model of neutrino masses \cite{PRL. 86 2502 (2001)} where
the smallness of the Seesaw neutrino mass may be naturally realized with $%
m_N $ (mass of right-handed singlet neutrinos) of order $1$ TeV. It is shown
that the Higgs mass of a new scalar doublet with lepton number $L=-1$ needed
in the model has to be larger than $50$ TeV to get the branching ratio of $%
\mu \rightarrow 3e$ to be consistant with the existing bound on $\mu
\rightarrow 3e$. This defeats the origional motivation of the model, namely
that there is no physics beyond the TeV\ energy scale.
\end{abstract}

\vspace{2in}

\begin{center}
{\Huge Acknowledgements}
\end{center}

\vspace{1.0cm} \thispagestyle{empty}

I am filled with the praise and glory to All Mighty Allah, the most merciful
and benevolent, who created the uneverse, with ideas of beauty, symmetry and
harmony, with regularity and without any chaos, and gave us the abilities to
discover what He thought.

Bless MUHAMMAD (P.B.U.H) the seal of the prophets and his pure and pious
progeny.

First and the foremost, I want to express my immense gratitude to my worthy
supervisor Prof. Dr. Riazuddin for his professsional guidance and
encouragement throughout this research work. Without his kind and invaluable
help it would have almost been impossible for me to accomplish this job. I
wish to say thanks to all the teachers at the Physics department because it
is due to them that we have an excellent environment of study in the
department. I can never forget the guidance of Dr. Fayyazuddin, Dr. Pervez
Hoodbhoy and Dr. Khurshaid Husnain throughout my studies.

My heartiest thanks to all members of the Particle Physics Group at QAU, and
especially to Shahzad, Irfan, Ijaz and Saba who helped me in one way or
other. I also wish to thank Mr. Amjad Hussain Shah Gilani for his
unforgetable help, cooperation and useful discussions which enable me to
hunt this task. I extend my thanks to Dr. Kashif Sabeeh and Dr. Hafeez
Hoorani for there encourging attitude.

My cordial thanks to my friends, Farhan, Moetasim, Assad, Hassan, Anees,
Adnan, Imran, Ansar, Irfan, Anis, Ali, Tasawar, Farooque, Jawad, Asia, Azra,
Aisha, Faiza, Zubair, Jamal, Anwar and Usman. There warm companionship make
the life at campus beautiful and unforgetable.

Finally I wish to record my deepest obligation to my parents, brothers and
sisters for their prayers, encouragment and finincial support during my
studies.

I also wish to acknowledge the financial support provided by Dr. Riazuddin
during my research work.\thispagestyle{empty}

\vspace{1.0in} {{{%
\begin{flushright}
{\bf \underline{Muhammad Jamil Aslam}}
\end{flushright}}}}

\newpage\ 

.\vspace{1.0cm} \thispagestyle{empty}

\begin{center}
This work is submitted as a dissertation

in partial fulfilment of

the requirement for the degree of

\vspace{1.0in}

{\bf MASTER OF PHILOSOPHY}

in

{\bf PHYSICS}

\vspace{1.0in}

Department of Physics

Quaid-e-Azam University

Islamabad 45320

Pakistan
\end{center}

\newpage\ 

.

\vspace{2.0in} \thispagestyle{empty}

\begin{center}
{\it TO MY LOVING {\bf MAA'N} Jee AND {\bf BABA} Jee}
\end{center}

\newpage\ 

{\Huge Certificate}

Certified that the work contained in this dissertation was carried out by
Mr. Muhammad Jamil Aslam under my supervision.

\vspace{1.0in} 
\begin{flushright}

(Prof. Dr. Riazuddin)

Supervisor

National Center for Physics

Quaid-e-Azam University

Islamabad 45320

\end{flushright}
\vspace{1.0cm} \thispagestyle{empty}

Submitted through:

\vspace{1.0cm}

(Prof. Dr. Asghari Masood)

Chairperson

Department of Physics

Quaid-e-Azam University

Islamabad 45320

\chapter{Introduction}

The most fundamental element of physics is the reduction principle. The
large variety of macroscopic forms of matter can be traced back according to
this principle, to a few microscopic constituents which interact by a small
number of forces. The reduction principle has provided a guide to unraveling
of the structure of physics from the macroscopic world through atomic and
nuclear physics to particle physics. The laws of nature are summarized in
the Standard Model of the particle physics.\setcounter{page}{1} %
\pagenumbering{arabic}

Standard Model is one of the successful model of the 20th century proposed
by the Glashow, Weinberg and Salam. The weak and electromagnetic
interactions are unified in the electroweak Standard Model. This model has
provided the plenty of successful predictions with an impressive level of
precision.

In quantum electrodynamics (QED) the interactions are specified by the gauge
principle. The electroweak Standard Model is based on the gauge symmetry
group $SU(2)\times U(1)$. $SU(2)$ is non-Abelian electroweak-isospin group,
to which three $W$ gauge fields are associated and $U(1)$ is the Abelian
hypercharge group. The associated $B$ field and the neutral component of the 
$W$ triplet field mix to form the photon field $A$ and a new neutral
electroweak field $Z$. This gauge group is spontaneously broken by the Higgs
mechanism (see for example 2.3.1). In the Standard Model, all particles
acquire their masses by interaction with another particle, the Higgs Boson.
Using the knowledge of basic symmetry group, the gauge invariant Lagrangian
has been written, giving not only the interactions of the various fields
including fermions but also the mass relationships for the fermions, gauge
bosons and Higgs boson (see for example 2.3.2).

Despite lot of success of the Standard Model, no body can say that it is the
end of physics. There are some limitations (see for example 2.4) which
necessiate the extension of the Standard Model.

When we advertize the Standard Model, we have said that it is a model whose
foundation is symmetry. In this model only those reactions are allowed which
conserve the individual as well as the total lepton-flavour. Thus the
reactions violating the lepton flavour can not be contained in the Standard
Model and cause its extension. This extension is commonly known as physics
beyond the Standard Model.

The problem of the physics beyond the Standard Model has been studied for a
considerable length of time. In the past few years substantial progress has
been made to understand this physics. Lepton Flavor Violating (LFV)
interactions are among the most promising candidates to understand this new
physics.

Using the experimental bounds on the three boby Lepton Flavor Violating
decays we have found the bounds on the two body LFV decays. These bounds are
suppressed by the form factors and was named as the dynamical suppression of
the LFV bounnds. The detailed study of these LFV\ decays is presented in
chapter 3. The bounds thus found on the two body LFV decays are used for
further studies of LFV decays in some interesting reactions.

Today there is evidence for a very important new property of the neutrinos
i.e. they have mass and, as a result mix with each other to lead the
phenomenon of neutrino osscillation. Evidences that the neutrinos are
massive particles come from three anomalous effects, the LSND excess, the
atmospheric anomaly and solar neutrino deficit. But the atmospheric and the
solar results are the most convincing one.The LSND has the small probability
compared to the atmospheric and solar anomalies. In order to incorporate
neutrino mass\ we must violate the Standard Model and introduce the new
concept of LFV. A Seesaw model of neutrino masses \cite{PRL. 86 2502 (2001)}
which involve right handed singlet neutrinos of mass $m_{N}$ of the order
TeV, is also disscussed using the LFV physics at loop order. The main
purpose of this model is that there is no new physics beyond $1$ TeV. It is
however, shown that even if one keeps $m_{N}$ at $1$ TeV, the present bounds
on $\mu \rightarrow 3e$ requires that new Higgs boson, necessary in this
model, to have mass larger than $50$ TeV. This necessiates a new physics
beyond the TeV energy scale defeating the orgional motivation of the model.
This is discussed in chapter 4, where it is also shown that the LFV
anamolous muou decay $\mu \rightarrow e\nu _{l}\bar{\nu}_{l}$ cannot
significantly contribute to the LSND DAR result, at least when we use $%
\tilde{g}_{Z_{\mu e}}$ as constrained by $\mu \rightarrow 3e$ in a model
independent way.

\chapter{The Electroweak Theory}

\section{History}

The proposal of the symmetry group for the Electroweak Theory, $SU(2)\times
U(1)$, was made by Glashow in 1961. His motivation was to unify weak and
electromagetic interactions into a symmetry group that contained $U(1)_{em}$%
. The prediction includes the existance of four physical vector boson
eigenstates, $W^{\pm },$ $Z,$ and $\gamma ,$ obtained from the rotations of
the weak eigenstate. In particular, the rotation by the weak angle $\theta
_{w}$ which defines the $Z$ weak boson was introduced already in this work.
Furthermore the correct structure of weak neutral current mediated by $Z$%
-boson was also obtained.The massive weak bosons $W^{\pm }$ and $Z$ were
considered as mediators of weak interactions. This model has serious problem
of giving masses to $W^{\pm }$ and $Z$, since the gauge symmetry would
predict their masses to be zero.The vector boson masses $M_{W}$ and $M_{Z}$
were parameters introduced by hand and the interaction Lagrangian was that
of the IVB Theory. The mass term for vector bosons in the Lagrangian not
only destroyed the gauge symmetry but also normalizability of the theory.

Another key ingredient for the building of the Electroweak Theory is
provided by the Goldstone Theorem which was introduced by Nambu in 1960 and
proved and studied with generality by Goldstone in 1961 and by Goldstone,
Salam and Weinberg in 1962. This theorem states the existance of massless
spinless particles as an implication of spontaneous symmetry breaking of
global symmetries.

The spontaneous symmetry breaking of local (gauge) symmetries, needed for
the breaking of the electroweak symmetry $SU(2)\times U(1)$, was studied by
P.Higgs, F. Englert and R. Brout, Guralnik, Hagen and Kibble in 1964. The
procedure for this spontaneous breakdown of gauge symmetries is reffered to
as the Higgs Mechanism.

The electroweak theory as it is now known was formulated by Weinberg and
independently by Salam in 1968 who incorporated the gauge group $SU(2)\times
U(1)$ introduced by the Glashow earlier. This theory, commonly called
Glashow-Weinberg Salam Model or Standard Model (SM), was built with the help
of the gauge principle and incorporated all the good phenomenological
properities of the pregauged theories of the weak interactions, and in
particular those of the IVB theory. It incorporated the idea of spontaneous
breaking of the gauge symmetry by introducing Higgs doublet. In this way the
weak vector bosons acquaired their masses with out destroying the
normalizability of the gauge theory.The SM is indeed a gauge theory of the
electroweak interactions based on the gauge symmetry $SU(2)\times U(1)$ and
the intermediate vector bosons, $\gamma ,W^{\pm }$ and $Z$ are the four
associated gauge bosons. The gauge boson masses, $M_{W}$ and $M_{Z}$, are
generated by the Higgs Mechanism in the Electroweak Theory and, as a
consequence, it respects unitarity at all energies and is renormalizable.

The important proof of renormlizability of gauge theories with and without
spontaneous symmetry was provided by 't Hooft in 1971.

The first firm indication that the Standard Model was the correct theory of
Electroweak interaction was the discovery of weak neutral Current in 1973 as
predicted by the model. This also provided the first measurement of $\sin
^2\theta _w$. By using this experimental input for $\theta _w$ and the
values of electromagnetic coupling and $G_F$, the SM provided the first
estimates for $M_W$ and $M_Z$ which were discovered experimently in 1983 at
the predicted masses. Another important ingredients of the SM are: fermion
family replication, quark mixing and CP violation.

The success of the SM was clearly the discovery of the gauge bosons $W^{\pm
} $ and $Z$ at the SpS collider at CERN in 1983. Since then there have been
plenty of tests of the SM even at quantum level \cite{9812242}.

\section{Choice of the group $SU(2)\times U(1)$}

In order to give an argument for the choice of $SU(2)\times U(1)$ gauge
group for electroweak unification, it is sufficient to consider the $%
e^{-}\nu $ component of the charged weak current that we write now in the
form, 
\begin{eqnarray*}
J_{\mu } &=&\bar{\nu}\gamma _{\mu }\left( \frac{1-\gamma _{5}}{2}\right) e=%
\bar{\nu}_{L}\gamma _{\mu }e_{L}=\bar{\psi}_{L}\gamma _{\mu }\tau _{+}\psi
_{L} \\
J_{\mu }^{\dagger } &=&\bar{e}\gamma _{\mu }\left( \frac{1-\gamma _{5}}{2}%
\right) \nu =\bar{e}_{L}\gamma _{\mu }\nu _{L}=\bar{\psi}_{L}\gamma _{\mu
}\tau _{-}\psi _{L}
\end{eqnarray*}
and we have introduced the lepton doublet notation and the $\tau _{i}$ ($%
i=1,2,3$) are Pauli matrices, 
\begin{eqnarray*}
\psi _{L} &=&\left( 
\begin{array}{c}
\nu _{L} \\ 
e_{L}
\end{array}
\right) ,~\bar{\psi}_{L}=\left( 
\begin{array}{cc}
\bar{\nu}_{L} & \bar{e}_{L}
\end{array}
\right) ,~\tau _{\pm }=\frac{1}{2}\left( \tau _{1}\pm i\tau _{2}\right) \\
\tau _{1} &=&\left( 
\begin{array}{cc}
0 & 1 \\ 
1 & 0
\end{array}
\right) ,~\tau _{2}=\left( 
\begin{array}{cc}
0 & -i \\ 
i & 0
\end{array}
\right) ,~\tau _{3}=\left( 
\begin{array}{cc}
1 & 0 \\ 
0 & -1
\end{array}
\right) .
\end{eqnarray*}
The three generators $I_{i}$ of unitary, unimodular group $SU(2)$ satisfy
the commutation relation 
\[
\left[ I_{i},I_{j}\right] =i\epsilon _{ijk}I_{k} 
\]
i.e. the same commutation relation as satisfied by $\tau _{i}$ viz 
\[
\left[ \tau _{i},\tau _{j}\right] =2i\epsilon _{ijk}\tau _{k}, 
\]
since for the fundamental representation of $SU(2)$ 
\[
I_{i}=\frac{\tau _{i}}{2} 
\]

Note that in the charged currents there are just two generators $I_1$ and $%
I_2$. A third generator $I_3$ is needed in order to close the $SU(2)$
algebera. This implies the third current that is relevent for the
electroweak interactions, 
\[
J_\mu ^3=\bar{\psi}_L\gamma _\mu \frac{\tau _3}2\psi _L=\frac 12\left( \bar{%
\nu}_L\gamma _\mu \nu _L-\bar{e}_L\gamma _\mu e_L\right) . 
\]

Obviously this can not be identified with the $J_\mu ^{em}$ which is $-\bar{e%
}\gamma _\mu e$. This clearly indicate that $SU(2)$ is not sufficient for
electroweak unification and it must be extended and the simple extension is
to consider the group $SU(2)\times U(1).$

The Gell-Mann Nishijima relation, 
\[
Q=I_3+\frac Y2 
\]
where,

$Q=$ electric charge, $I_{3}=$ weak isospin, $Y=$ weak hypercharge; implies 
\begin{eqnarray*}
Y &=&-1\text{\hspace{1in} for }\nu _{L} \\
Y &=&-1\text{\hspace{1in} for }e_{L} \\
Y &=&-2\text{\hspace{1in} for }e_{R}
\end{eqnarray*}
The corresponding relation amoung the current is 
\[
J_{\mu }^{em}=J_{\mu }^{3}+\frac{1}{2}J_{\mu }^{Y} 
\]
Thus 
\[
J_{\mu }^{Y}=2\left( J_{\mu }^{em}-J_{\mu }^{3}\right) 
\]

Therefore, if the following are used as inputs 
\begin{eqnarray*}
J_{\mu }^{em} &=&(-1)\bar{e}_{L}\gamma _{\mu }e_{L}+(-1)\bar{e}_{R}\gamma
_{\mu }e_{R} \\
J_{\mu }^{3} &=&(-\frac{1}{2})\bar{e}_{L}\gamma _{\mu }e_{L}+(\frac{1}{2})%
\bar{\nu}_{L}\gamma _{\mu }\nu _{L}
\end{eqnarray*}
one can get $J_{\mu }^{Y}$ as output, 
\[
J_{\mu }^{Y}=2\left( J_{\mu }^{em}-J_{\mu }^{3}\right) =(-1)\bar{e}%
_{L}\gamma _{\mu }e_{L}+(-2)\bar{e}_{R}\gamma _{\mu }e_{R}+(-1)\bar{\nu}%
_{L}\gamma _{\mu }\nu _{L}. 
\]
This clearly indicates that 
\begin{eqnarray*}
Y &=&-1\text{\hspace{1in} for }\left( 
\begin{array}{c}
e_{L} \\ 
\nu _{L}
\end{array}
\right) \\
Y &=&-2\text{\hspace{1in}\ for \quad }e_{R}
\end{eqnarray*}
Due to symmetry breaking the two neutral currents $J_{\mu }^{3}$ and $J_{\mu
}^{em}$ will mix to give two physical currents of which one must be
identified with electromagnetic current $J_{\mu }^{em}$ and the second
current will be new currents. These currents will be coupled to physical
vector bosons $A_{\mu }$ and $Z_{\mu }$. 
\begin{equation}
g_{2}J_{\mu }^{3}W_{3\mu }+\frac{1}{2}\acute{g}J_{\mu }^{Y}B_{\mu }=eJ_{\mu
}^{em}A_{\mu }+g_{Z}J_{\mu }^{Z}Z_{\mu }  \label{a1}
\end{equation}
where, 
\begin{eqnarray*}
A_{\mu } &=&\cos \theta _{w}B_{\mu }+\sin \theta _{w}W_{3\mu }, \\
Z_{\mu } &=&\cos \theta _{w}W_{3\mu }-\sin \theta _{w}B_{\mu };\theta _{w}=%
\text{ weak angle.}
\end{eqnarray*}
From here 
\begin{eqnarray*}
W_{3\mu } &=&\sin \theta _{w}A_{\mu }+\cos \theta _{w}Z_{\mu }, \\
B_{\mu } &=&\cos \theta _{w}A_{\mu }-\sin \theta _{w}Z_{\mu }.
\end{eqnarray*}
Thus, Eq. (\ref{a1}) becomes 
\begin{eqnarray*}
g_{2}J_{\mu }^{3}\left[ \sin \theta _{w}A_{\mu }+\cos \theta _{w}Z_{\mu
}\right] +\frac{1}{2}\acute{g}J_{\mu }^{Y}\left[ \cos \theta _{w}A_{\mu
}-\sin \theta _{w}Z_{\mu }\right] &=&eJ_{\mu }^{em}A_{\mu }+g_{Z}J_{\mu
}^{Z}Z_{\mu }. \\
&& \\
&&
\end{eqnarray*}
So, 
\begin{eqnarray*}
eJ_{\mu }^{em} &=&g_{2}J_{\mu }^{3}\sin \theta _{w}+\frac{1}{2}\acute{g}\cos
\theta _{w}J_{\mu }^{Y}=-e\left[ \bar{e}_{L}\gamma _{\mu }e_{L}+\bar{e}%
_{R}\gamma _{\mu }e_{R}\right] \\
&=&\frac{1}{2}g_{2}\sin \theta _{w}\left[ -\bar{e}_{L}\gamma _{\mu }e_{L}+%
\bar{\nu}_{L}\gamma _{\mu }\nu _{L}\right] \\
&&+\frac{1}{2}\acute{g}\cos \theta _{w}\left[ (-1)\bar{e}_{L}\gamma _{\mu
}e_{L}+(-2)\bar{e}_{R}\gamma _{\mu }e_{R}+(-1)\bar{\nu}_{L}\gamma _{\mu }\nu
_{L}\right] .
\end{eqnarray*}
This simply gives us 
\begin{eqnarray*}
g_{2}\sin \theta _{w} &=&e \\
\acute{g}\cos \theta _{w} &=&e
\end{eqnarray*}
Hence 
\[
\tan \theta _{w}=\frac{g_{2}}{\acute{g}}. 
\]
Similarly 
\begin{eqnarray*}
g_{Z}J_{\mu }^{Z} &=&g_{2}J_{\mu }^{3}\cos \theta _{w}-\frac{1}{2}\acute{g}%
J_{\mu }^{Y}\sin \theta _{w} \\
&& \\
&=&\frac{g_{2}}{\cos \theta _{w}}\left[ -\bar{e}_{L}\gamma _{\mu
}e_{L}\left( \frac{1}{2}-\sin ^{2}\theta _{w}\right) +\frac{1}{2}\bar{\nu}%
_{L}\gamma _{\mu }\nu _{L}+\sin ^{2}\theta _{w}\bar{e}_{R}\gamma _{\mu
}e_{R}\right] .
\end{eqnarray*}
This will gives us the neutral current $J_{\mu }^{Z}=J_{\mu }^{NC}$and the
corresponding coupling $g_{Z}$. This is the main indication of the
electroweak unification.

\section{The Electroweak Standard Model}

The Electroweak Standard Model is the commonly accepted theory of the
fundamental electroweak interactions \cite{ictp book}. When we want to
advertise the virtue of the Standard Model, we say that it is a model whoes
foundation is symmetry \cite{paskin}. It is a gauge invariant Quantum Field
Theory based on the symmetry group $SU(2)\times U(1)$, which is
spontaneously broken by the Higgs mechanism.

The Electroweak Standard Model consists of three components.

1): The basic constituents of matter are leptons and quarks which are
realized in three families of identical structure: 
\[
\begin{tabular}{|c|c|}
\hline
$\text{Leptons}$ & $
\begin{tabular}{|c|c|c|}
\hline
$\nu _{e}$ & $\nu _{\mu }$ & $\nu _{\tau }$ \\ \hline
$e^{-}$ & $\mu ^{-}$ & $\tau ^{-}$ \\ \hline
\end{tabular}
$ \\ \hline
$\text{Quarks}$ & 
\begin{tabular}{|l|l|l|}
\hline
$u$ & $c$ & $t$ \\ \hline
$d$ & $s$ & $b$ \\ \hline
\end{tabular}
\\ \hline
\end{tabular}
\]
We will concentrate on the leptonic sector only.

2): Four different forces act between leptons and quarks: The
electromagnetic and weak forces are unified in the Standard Model. The
fields associated with these forces are spin $1$ fields, describing the
photon $\gamma $ and the electroweak gauge bosons $W^{\pm }$ and $Z$.

3): The third component of the Standard Model is the Higgs mechanism.

Before going to the deep discussion of the Standard Model, we have to
require some theoretical basis \cite{SM and basis}.

\subsection{The Theoretical Base}

The fundamental forces of the Electroweak Standard Model, the
electromagnetic and the weak force, are mediated by gauge fields. The
concept could consistantly be extended to massive gauge field by introducing
the Higgs mechanism which generates masses with out destroying the
underlying gauge symmetries of the theory \cite{Higgs hunter}.

\subsubsection{\bf 1): Gauge Sector}

Gauge invariant theories are invariant under gauge transformations of
fermion fields: $\psi \rightarrow U\psi $. $U$ is either a phase factor for
Abelian transformations or Unitary matrix for non-Abelian transformation
acting on the multiplets of the fermion field $\psi $. Now if the theory
guarantee the local transformation for which $U$ depends on the space time
point $x$, the usual space-time derivatives $\partial _\mu $ must be
extended to covariant derivatives $D_{_\mu }$ which includes a new vector
field $V_\mu $: 
\[
i\partial _\mu \rightarrow iD_{_\mu }=i\partial _\mu -gV_\mu 
\]
$g$ defines the universal gauge coupling of the system. Under local gauge
transformations the gauge field $V_\mu $ is transformed by a rotation plus a
shift: 
\[
V_\mu \rightarrow UV_\mu U^{-1}+ig^{-1}\left[ \partial _\mu U\right] U^{-1}. 
\]

But in contrast to this, the curl $F$ of $V_\mu $, 
\[
F_{\mu \nu }=-ig^{-1}\left[ D_\mu ,D_\nu \right] 
\]
is just rotated under gauge transformation.

The Lagrangian describing the systen of spin$\frac 12$ fermions and
vectorial gauge bosons for massless particles can be written in the compact
form as follows: 
\[
L\left[ \psi ,V\right] =\bar{\psi}iD\psi -\frac 12TrF^2 
\]
It incorporates the following interactions:

Fermion-gauge bosons 
\[
-g\bar{\psi}V\psi . 
\]

Three bosons couplings 
\[
igTr\left( \partial _\nu V_{_\mu }-\partial _\mu V_\nu \right) \left[
V_{_\mu },V_{_\nu }\right] . 
\]

Four boson couplings 
\[
\frac 12g^2Tr\left[ V_{_\mu },V_{_\nu }\right] ^2. 
\]

\subsubsection{\bf 2): Higgs mechanism}

What is called the Higgs mechanism is the extension of the spontaneous
symmetry breaking to creat massive vector bosons in a gauge invariant
theory. As the SM is a gauge theory, the $SU(2)\times U(1)$ gauge invariance
requires masses of the gauge bosons to be zero, since the presence of an
explicit mass term for the gauge bosons in the Lagrangian violates gauge
invariance. The Higgs mechanism circumvents this constraint by begining with
a gauge invariat theory having massless gauge bosons.The $W^{\pm }$ and $%
Z^{0}$ masses were generated by spontaneously breaking the local gauge
symmetry from $SU(2)\times U(1)\rightarrow U(1)_{em}$, which was achieved by
the introduction of a self-interacting complex scalar field, $\Phi $,
transforming as an $SU(2)$ doublet. The doublet field $\Phi $ and its
complex conjugate togather comprise four independent fields. Spontaneous
symmetry breaking was implemented by giving one of the neutral fields a
nonzero vacuum expectation value, 
\[
\left\langle \phi \right\rangle \equiv \left\langle 0\left| \phi \right|
0\right\rangle =\frac{v}{\sqrt{2}}\neq 0. 
\]
Of the four fields in the Lagrangian before spontaneous symmetry breaking,
three fields become the longitudinal degrees of freedom of the vector bosons 
$W^{\pm }$and $Z^{0}$; the photon coupled to the remaining symmetry group $%
U(1)_{em}$-generators, remains massless.

One neutral scalar particle remains in the physical sector of the theory.
This is the so-called Salam-Weinberg Higgs particle, which the $SU(2)\times
U(1)$ model predicts to exist.

Since the same Higgs doublet is used to give masses to the bosons and
fermions, which have Yukawa couplings with the scalar fields, the $%
SU(2)\times U(1)$ model predicts the couplings of the Higgs particles with
all the known bosons and fermions but makes no prediction about its mass.
This could be traced back to the fact that in Salam-Weinberg theory, the
Higgs particle mass is function of the unknown quartic Higgs-boson coupling
constant.

\subsection{Formulation of the Electroweak Standard Model}

\subsubsection{\bf The Matter Sector}

The fundamental fermions, as families with left handed isospin doublets and
right handed isospin singlets appear in the fundemental representation of
the the group $SU(2)\times U(1)$. It is realized that the symmetry pattern
in the first, second and third generation of the fermions is same, 
\[
\left[ 
\begin{array}{c}
\nu _e \\ 
e^{-}
\end{array}
\right] _L 
\begin{array}{c}
\nu _{eR} \\ 
e_R^{-}
\end{array}
;\left[ 
\begin{array}{c}
\nu _\mu \\ 
\mu ^{-}
\end{array}
\right] _L 
\begin{array}{c}
\nu _{\mu R} \\ 
\mu _R^{-}
\end{array}
;\left[ 
\begin{array}{c}
\nu _\tau \\ 
\tau ^{-}
\end{array}
\right] _L 
\begin{array}{c}
\nu _{\tau R} \\ 
\tau _R^{-}
\end{array}
\]
The symmetry structure cannot be derived in the Standard Model. It is an
experimental fact that in weak interactions the parity is not conserved. The
different isospin assigned to the left handed and right handed field allows
for maximal parity violation in the weak interactions. So the experimental
observation is incorporated in the natural way.

The relationship between the electric charge $Q$ and basic quantum numbers
is described by Gell-Mann-Nishijima relation 
\[
Q=I_3+\frac Y2. 
\]

\subsubsection{\bf Interactions}

The interactions of the Standard Model are summarized by the three terms in
the basic Lagrangian: 
\begin{equation}
L=L_G+L_F+L_H  \label{201}
\end{equation}
which are specified in the following way;

\subsubsection{\bf Gauge fields}

$SU(2)\times U(1)$ is a non-Abelian group which is generated by the isospin
operators $I_{1}$, $I_{2}$, $I_{3}$ and the hypercharge $Y$. Each of these
generlized charges is associated with a vector field: a triplet of vector
fields $W_{\mu }^{1,2,3}$ with $I_{1,2,3}$ and a singlet field $B_{\mu }$
with $Y$. The isotriplet $W_{\mu }^{a}$, $a=1,2,3$ and isosinglet $B_{\mu }$
lead to the field strength tensors 
\begin{eqnarray}
W_{\mu \nu }^{a} &=&\partial _{\mu }W_{\nu }^{a}+\partial _{\nu }W_{\mu
}^{a}+g_{2}\varepsilon _{abc}W_{\mu }^{b}W_{\nu }^{c}  \nonumber  \label{2}
\\
B_{\mu \nu } &=&\partial _{\mu }B_{\nu }-\partial _{\nu }B_{\mu }
\label{202}
\end{eqnarray}
$g_{2}$ is defined as the coupling constant for non-Abelian gauge group $%
SU(2)$

Using the equation (\ref{202}) the pure gauge field Langragian can be
written as follows 
\begin{equation}
L_G=-\frac 14W_{\mu \nu }^aW^{\mu \nu ,a}-\frac 14B_{\mu \nu }B^{\mu \nu }
\label{203}
\end{equation}
It is invariant under the non-Abelian $SU(2)\times U(1)$ transformation.

\subsubsection{\bf Fermion fields and fermion-gauge interactions}

The left-handed fermion fields of each lepton family 
\[
\psi _j^L=\left( 
\begin{array}{c}
\psi _{j+}^L \\ 
\psi _{j-}^L
\end{array}
\right) 
\]
with family index $j$ are grouped into $SU(2)$ doublets with component index 
$\sigma =\pm $, and the right-handed fields into singlets 
\[
\psi _j^R=\psi _{j\sigma }^R. 
\]
Each left and right-handed multiplet is an eigenstate of the weak
hypercharge $Y$ such that the relation (\ref{202}) is fulfilled. The
Covariant derivative 
\begin{equation}
D_\mu =\partial _\mu -ig_2I_aW_\mu ^a+ig_1\frac Y2B_\mu  \label{204}
\end{equation}
induces the fermion-gauge field interaction via the minimal subsitution rule 
\begin{equation}
L_F=\sum_j\bar{\psi}_j^Li\gamma ^\mu D_\mu \psi _j^L+\sum_{j,\sigma }\bar{%
\psi}_{j\sigma }^Ri\gamma ^\mu D_\mu \psi _{j\sigma }^R.  \label{205}
\end{equation}
$g_1$ is the coupling constant for the Abelian $U(1)$ gauge group.

\subsubsection{\bf Higgs field and Higgs interaction}

For spontaneous breaking of the $SU(2)\times U(1)$ symmetry leaving the
electromagnetic gauge group $U(1)$ unbroken, a single complex scalar doublet
field with hypercharge $Y=1$%
\begin{equation}
\phi \left( x\right) =\left( 
\begin{array}{c}
\phi ^{+}\left( x\right) \\ 
\phi ^0\left( x\right)
\end{array}
\right)  \label{206}
\end{equation}
is coupled to the gauge fields through 
\begin{equation}
L_H=\left( D_\mu \phi \right) ^{+}\left( D^\mu \phi \right) -V\left( \phi
\right)  \label{207}
\end{equation}
with the covariant derivative 
\[
D_\mu =\partial _\mu -ig_2I_aW_\mu ^a+ig_1\frac{B_\mu }2. 
\]
The Higgs field self interaction 
\begin{equation}
V\left( \phi \right) =-\mu ^2\phi ^{+}\phi +\frac \lambda 4\left( \phi
^{+}\phi \right) ^2  \label{208}
\end{equation}
is constructed in such a way that $\phi $ has a non vanishing vacuum
expectation value, i.e. 
\[
\left\langle \phi \right\rangle =\frac 1{\sqrt{2}}\left( 
\begin{array}{c}
0 \\ 
v
\end{array}
\right) 
\]
with 
\begin{equation}
v=\frac{2\mu }{\sqrt{\lambda }}.  \label{209}
\end{equation}

The field (\ref{206}) can be written in the following way, 
\begin{equation}
\phi \left( x\right) =\left( 
\begin{array}{c}
\phi ^{+}\left( x\right) \\ 
\left( v+H\left( x\right) +i\chi \left( x\right) \right) /\sqrt{2}
\end{array}
\right)  \label{210}
\end{equation}
where the field components $\phi ^{+}$, $H$, $\chi $ have vacuum expectation
values zero.

Exploiting the invariance of the Langragian, the components $\phi ^{+}$, $%
\chi $ can be gauged away; this means that they are unphysical (Higgs ghosts
or would be Goldstone bosons). In this particular gauge, the unitarity
gauge, the Higgs field has simple form 
\[
\phi \left( x\right) =\frac 1{\sqrt{2}}\left( 
\begin{array}{c}
0 \\ 
v+H\left( x\right)
\end{array}
\right) . 
\]
The real field $H\left( x\right) $ which describes small osscillations about
the ground state defines the physical Higgs field.

The Higgs field components have triplet and quartic self couplings following
from $V$ and couplings to the gauge fields via the kinetic term of Eq.(\ref
{207}).

In addition, Yukawa couplings of the fermions are introduced in order to
make the fermion massive. The Lagrangian for the Yukawa term can be written
as follows 
\begin{equation}
L_{Yukawa}=g_l\left( \nu _L\phi ^{+}l_R+\bar{l}_R\phi ^{-}\nu _L+\bar{l}%
_L\phi ^0l_R+\bar{l}_R\phi ^{0*}l_L\right)  \label{211}
\end{equation}
The fermion mass terms follow from the $v$- part of $\phi ^0$ \cite{ictp
book}.

The Lagrangian $L$ summarizes the laws of physics for the electromagnetic
and the weak interactions between the leptons, and it predicts the form of
self-interaction between the gauge fields. Morover, the specific form of the
Higgs interaction generates the mass of the fundamental particles, the
leptons, the gauge bosons and the Higgs boson itself, and it predicts the
interactions of the Higgs particle \cite{SM and basis}.

\subsubsection{\bf Masses and mass eigenstates of particles}

In the unitary gauge the mass terms are extracted by substiuting 
\[
\left[ \phi \rightarrow 0,\frac v{\sqrt{2}}\right] 
\]
in the basic Higgs Lagrangian (\ref{207}).\ The apparent $SU(2)$ seems to be
lost thereby, but only superficially so and remain present in the hidden
form; the resulting Lagrangian preserves an apparent local $U\left( 1\right) 
$ gauge symmetry which is identified with the electromagnetic gauge
symmetry: $SU(2)\times U\left( 1\right) \rightarrow U\left( 1\right) _{em}$ 
\cite{SM and basis}.

\subsubsection{\bf Gauge Bosons}

The mass matrix of the gauge boson in the basis $\left( \vec{W},B\right) $
takes the form 
\begin{equation}
M_{V}^{2}=\frac{1}{4}v^{2}\left( 
\begin{array}{cccc}
g_{W}^{2} &  &  &  \\ 
& g_{W}^{2} &  &  \\ 
&  & g_{W}^{2} & g_{W}\acute{g}_{W} \\ 
&  & g_{W}\acute{g}_{W} & \acute{g}_{W}^{2}
\end{array}
\right) .
\end{equation}
This gives the mass of the vector boson in non diagonal form. The mass of
the charged weak bosons is obvious 
\[
M_{W^{\pm }}^{2}=\frac{1}{4}g_{W}^{2}v^{2}. 
\]
As eigenstates related to the two masses $M_{W^{\pm }}^{2}$ the charged $%
W^{\pm }$ boson state may be defined as 
\begin{equation}
W_{\mu }^{\pm }=\frac{1}{\sqrt{2}}\left[ W_{\mu }^{1}\mp W_{\mu }^{2}\right]
.
\end{equation}

For the neutral bosons, the mass term gives the matrix 
\begin{equation}
M_{V_{N}}^{2}=\frac{1}{4}\left( 
\begin{array}{cc}
g_{W}^{2} & g_{W}\acute{g}_{W} \\ 
g_{W}\acute{g}_{W} & \acute{g}_{W}^{2}
\end{array}
\right) v^{2}
\end{equation}
Since $\det (M_{V_{N}}^{2})=0$, therefore one of the eigenvalue of $%
M_{V_{N}}^{2}$ is zero. The above matrix is diagonalized by defining the
fields $A_{\mu }$, $Z_{\mu }$: 
\begin{eqnarray}
A_{\mu } &=&\cos \theta _{W}B_{\mu }+\sin \theta _{W}W_{\mu }^{3} \\
Z_{\mu } &=&-\sin \theta _{W}B_{\mu }+\cos \theta _{W}W_{\mu }^{3}
\end{eqnarray}
In matrix form the above equations can be written as follows; 
\begin{equation}
\left( 
\begin{array}{c}
A_{\mu } \\ 
Z_{\mu }
\end{array}
\right) =\left( 
\begin{array}{cc}
\cos \theta _{W} & \sin \theta _{W} \\ 
-\sin \theta _{W} & \cos \theta _{W}
\end{array}
\right) \left( 
\begin{array}{c}
B_{\mu } \\ 
W_{\mu }^{3}
\end{array}
\right)
\end{equation}
Then we get 
\begin{eqnarray}
M_{A}^{2} &=&0\qquad A_{\mu }:~\text{photon} \\
M_{Z}^{2} &=&\frac{1}{4}\left( g_{W}^{2}+\acute{g}_{W}^{2}\right) v^{2} 
\nonumber \\
&=&\frac{1}{4}g_{W}^{2}v^{2}\left( \frac{1}{\cos ^{2}\theta _{W}}\right)
\end{eqnarray}
Where 
\begin{equation}
\tan \theta _{W}=\frac{\acute{g}_{W}}{g_{W}}
\end{equation}
i.e. the electroweak mixing angle $\theta _{W}$ is defined by the ratio of
the $SU(2)$ and $U(1)$ couplings.

Introduce a parameter 
\[
\rho =\frac{M_W^2}{M_Z^2\cos ^2\theta _W} 
\]
Now using the value of $M_Z^2$, we get 
\[
\rho =1 
\]
This is the consquence of the fact that Higgs field is a doublet under $%
SU(2)_L$ \cite{FayyazuddinandRiazudin}.

Experimently the mixing angle turns out to be large, i.e. $\sin ^2\theta
_W\simeq 0.23$. The fact that the experimental value for $\sin ^2\theta _W$
is far away from the limits $0$, or $1$, indicate a large mixing effect.
This supports the interpretation that the electromagnetic and the weak
interactions are indeed manifestations of a unified electroweak interaction
even though the underlying symmetry group $SU(2)\times U\left( 1\right) $ is
not simple. It may therefore be concluded that the electromagnetic and the
weak interactions are truly unified in the Glashow-Salam-Weinberg theory of
the electroweak interactions.

The ground-state value of the Higgs field is releated to the Fermi coupling
constant. From the low-energy relation $G_F/\sqrt{2}=g_W^2/8M_W^2$ in $\beta 
$ decay and combine with the mass relation $M_{W^{\pm }}^2=\frac 14g_W^2v^2$%
, the value of $v$ can be derived: 
\begin{eqnarray}
v &=&[1/\sqrt{2}G_F]^{1/2}  \nonumber \\
&\simeq &246GeV
\end{eqnarray}
The typical range for electroweak phenomena, defined by the weak masses $M_W$
and $M_Z$, is of the order $100GeV$.

\subsubsection{\bf Fermions}

The leptons are endowed with mass by means of Yukawa interactions with the
Higgs ground state: 
\begin{equation}
M_f=g_f\frac v{\sqrt{2}}.
\end{equation}
Though the masses of chiral fermion fields can be introducd in a consistant
way via the Higgs mechanism, the Standard Model does not provide predictions
for the experimental values of the Yukawa couplings $g_f$ and, as
consquence, of the mases. The theory of the masses is not available yet.

\subsubsection{\bf The Higgs Bosons}

The real field $H\left( x\right) $ which describes small osscillations about
the ground state tells us the mass of the physical neutral scalar particles
with mass 
\begin{equation}
M_H=\mu \sqrt{2}=\sqrt{\lambda }v.
\end{equation}
It can not predicted in the Standard Model since the quartric coupling $%
\lambda $ is an unknown parameter.

We conclude this session with the folowing remarks:

1). A definite prediction of electroweak unification is the existance of
weak neutral currents with the same effective couplings as charged currents.
This current has been found experimentaly.

2). The existance of the vector bosons $W^{\pm }$, $Z$ with definite masses
which have also been discovered.

3). The theory has one free parameter $\sin ^2\theta _W.$

\section{Limitations of the Standard Model}

There is no confirmed experimental evidence from accelerators against the
Standard Model, and several possible extensions have been ruled out.
Nevertheless, there is no thinking physicst could imagine that the Standard
Model is the end of physics. Even if one accepts the strange set of group
representations and hypercharge that it requires, the Standard Model
contains at least $19$ parameters. Moreover, many more parameters are
required if one wishes to accommodate non-accelerator observations. For
example, the neutrino masses and mixing introduce at least $7$ parameters: $%
3 $ masses, $3$ mixing angles and $1$ CP violating phase \cite{9812235}.

\section{Summary}

In this chapter, the $SU(2)$ and $U(1)$ gauge group have been reviewed,
which provides a back ground knowledge for the understanding of the
Electroweak Standard Model. The concept of how vector bosons $W^{\pm }$, $Z$
acquire masses have also been discussed in this chapter. The neutrino masses
and mixings which are the weaknesses of the Standard Model have been
mentioned here and will be studied in detail in the last chapter.


\chapter{Generic Feature of Lepton Flavour Violation}

\section{Introduction}

It is generally believed that the Standard Model of electroweak interactions
is a low-energy approximation to a more fundamental theory. Yet there is no
clear experimental evidence either to guide its extension to additional
physical processes or to predict the model parameters. The Standard Model
incorporates the lepton family-number conservation, which has been
empirically verified to high precision but is not a consequence of a known
gauge theory . Indeed many theoretical extensions to the Standard Model
allow lepton family-number violation within a range that can be tested by
experiment \cite{MEGA Collaboration}.

The predications of the rate for a given family-number nonconserving process
vary among these extensions, and thus provide a test of the model. Many
possibilities have been explored, and the present experimental limits for a
wide variety of the processes have been tabulated. Thus the Lepton Flavour
Violation (LFV) processes that are strongly suppressed in the Standard Model
by powers of (small) neutrino masses may provide signals for new physics. At
present we have strigent bound for $\mu $ decays, e.g. 
\[
BR\left( \mu \rightarrow 3e\right) \leq 10^{-12} 
\]
and some what weaker $O(10^{-16})$ bounds on LFV $\tau $ decays \cite{S.
Nussinov}.

The possibility of large samples of decaying vector bosons [$V=J/\psi
,\Upsilon ,$ and $Z^0$] and the clear signature provided by $\mu ^{\pm
}e^{\pm }$ final states suggest searching for LFV two-body decays 
\[
V\rightarrow \mu ^{\pm }e^{\pm }. 
\]
Here we show that rather simple considerations based on unitarity, provide
rather strong constraints on two-body LFV processes. Hence, most three-body $%
\mu $ and $\tau $ LFV decays are likely to provide more sensitive tests of
lepton flavour violation, rather than the two-body decays.

\section{Basic consideration and calculation}

Let us assume that a vector boson $V_{i}$ (here $V_{i}$ could be either a
fundamental state, such as the $Z^{0}$ or the quark- antiquark bound state
such as the $\phi ,J/\psi ,$ or $\Upsilon $) couples to $\mu ^{\pm }e^{\pm }$%
. If it couples also to $e^{+}e^{-}$, as all the states above do, then by
unitarity $V$ exchange contributes also to $\mu \rightarrow 3e$. Let us
define the effctive couplings between vector boson $V_{i}$ and $\mu ^{\pm
}e^{\pm }$ as $\tilde{g}_{V_{\mu e}}$, and the corresponding effective
Lagrangian can be written as 
\begin{equation}
{\cal L}_{eff}=\tilde{g}_{V_{\mu e}}\bar{\mu}\gamma ^{\alpha }eV^{\alpha
}+H.c.  \label{301}
\end{equation}
From the Feynman diagram, the amplitude $A\left( \mu \rightarrow 3e\right) $
can be written as 
\begin{eqnarray}
A\left( \mu \rightarrow 3e\right)  &=&\bar{u}_{e}(k_{3})\tilde{g}_{V_{\mu
e}}\gamma ^{\alpha }u_{\mu }(p)\frac{g_{\alpha \beta }}{M_{V}^{2}-s}\bar{u}%
_{e}(k_{2})g_{V_{ee}}\gamma ^{\beta }v_{e}(k_{1})  \nonumber \\
&=&\bar{u}_{e}(k_{3})\gamma ^{\alpha }u_{\mu }(p)\bar{u}_{e}(k_{2})\gamma
_{\alpha }v_{e}(k_{1})\frac{\tilde{g}_{V_{\mu e}}g_{V_{ee}}}{M_{V}^{2}-s}.
\label{302}
\end{eqnarray}
Here $g_{V_{ee}}$ is the effective coupling of the vector boson $V_{i}$ to $%
e^{+}e^{-}$, while 
\[
s\equiv (k_{1}+k_{2})^{2}\leq m_{\mu }^{2}.
\]
(There are, of course, also potential axial vector couplings of $V$ to $%
e^{+}e^{-}$, which contribute to this process. These can be included in the
above, but as we shall see they do not change qualitative discussion).

Since we are dealing with the low energy process, as a first approximation,
it is sensible to neglect $s$ in comparision with $M_V^2$. Therefore the Eq.
(\ref{302}) takes the form, 
\begin{equation}
A(\mu \longrightarrow 3e)=\bar{u}_e(k_3)\gamma ^\alpha u_\mu (p)\bar{u}%
_e(k_2)\gamma _\alpha v_e(k_1)\frac{\tilde{g}_{V_{\mu e}}g_{V_{ee}}}{M_V^2}.
\label{303}
\end{equation}
In order to calculate the decay width we have to calculate $\left| A\right|
^2$ which is given by 
\begin{eqnarray}
\left| A\right| ^2 &=&AA^{\dagger }  \nonumber \\
&=&\bar{u}_e(k_3)\gamma ^\alpha u_\mu (p)\bar{u}_e(k_2)\gamma _\alpha
v_e(k_1)\bar{v}_e(k_1)  \nonumber  \label{06} \\
&&\times \gamma _\rho u_e\left( k_2\right) \bar{u}_\mu (p)\gamma ^\rho
u_{_e}\left( k_3\right) \left( \frac{\tilde{g}_{V_{\mu e}}g_{V_{ee}}}{M_V^2}%
\right) ^2.  \label{306}
\end{eqnarray}
Taking the spin average for the initial state and summation over spins for
final state particles, we get 
\begin{eqnarray}
\sum_{spins=\pm \frac 12}\left| A\right| ^2 &=&\frac 1{16}\left[ Tr\left[
\gamma _\alpha \left( \not{k}_1-m_e\right) \gamma _\rho \left( \not%
{k}_2+m_e\right) \right] \right.  \nonumber   \\
&&\left. \times Tr\left[ \gamma ^\alpha (\not{p}+m_{_\mu })\gamma ^\rho
\left( \not{k}_3+m_e\right) \right] \right] \left( \frac{\tilde{g}_{V_{\mu
e}}g_{V_{ee}}}{M_V^2}\right) ^2.  \label{309}
\end{eqnarray}
Assuming that the external particles (electrons) are massless. Therefore,
Eq. (\ref{309}) takes the form 
\begin{equation}
\sum_{spins=\pm \frac 12}\left| A\right| ^2=2\left[ \left( k_1\cdot p\right)
\left( k_2\cdot k_3\right) +\left( k_1\cdot k_3\right) \left( k_2\cdot
p\right) \right] .  \label{314}
\end{equation}

We next carry the phase space integrations, starting with the integrals over
the electrons momenta, given by 
\begin{equation}
I^{\mu \nu }\left( q\right) =\int d^{3}{\bf k}_{{\bf 1}}d^{3}{\bf k}_{{\bf 2}%
}\frac{k_{1}^{\mu }k_{2}^{\nu }}{E_{1}E_{2}}\delta ^{4}\left(
k_{1}+k_{2}-q\right)  \label{315}
\end{equation}
where 
\begin{equation}
q\equiv p-k_{3}.  \label{316}
\end{equation}
It follows from the Lorentz covariance of the integral (\ref{315}) that the
most general form is 
\begin{equation}
I^{\mu \nu }\left( q\right) =g^{\mu \nu }A\left( q^{2}\right) +q^{\mu
}q^{\nu }B\left( q^{2}\right) .  \label{317}
\end{equation}
From this equation it follows that 
\begin{eqnarray}
g_{\mu \nu }I^{\mu \nu }\left( q\right) &=&4A\left( q^{2}\right)
+q^{2}B\left( q^{2}\right)  \nonumber \\
q_{\mu }q_{\nu }I^{\mu \nu }\left( q\right) &=&q^{2}A\left( q^{2}\right)
+\left( q^{2}\right) ^{2}B\left( q^{2}\right) .  \label{318}
\end{eqnarray}
This can be obtained by using Eq. (\ref{317}).

Since we have been taking the electron masses to be zero so that $%
k_1^2=k_2^2=k_3^2=0$ and, on account of the $\delta $-function in (\ref{315}%
), 
\begin{eqnarray}
q^2 &=&\left( k_1+k_2\right) ^2=2\left( k_1\cdot k_2\right)  \nonumber \\
\frac{q^2}2 &=&k_1\cdot k_2  \label{319}
\end{eqnarray}

In order to find $A\left( q^{2}\right) $ and $B\left( q^{2}\right) $, we
calculate the expersion on the left hand sides of (\ref{318}). From Eq. (\ref
{315}) we obtain 
\begin{eqnarray}
g_{\mu \nu }I^{\mu \nu }\left( q\right) &=&\int d^{3}{\bf k}_{{\bf 1}}d^{3}%
{\bf k}_{{\bf 2}}\frac{g_{\mu \nu }k_{1}^{\mu }k_{2}^{\nu }}{E_{1}E_{2}}%
\delta ^{4}\left( k_{1}+k_{2}-q\right)  \nonumber \\
&=&\int d^{3}{\bf k}_{{\bf 1}}d^{3}{\bf k}_{{\bf 2}}\frac{\left(
k_{1}.k_{2}\right) }{E_{1}E_{2}}\delta ^{4}\left( k_{1}+k_{2}-q\right) .
\label{320}
\end{eqnarray}
Using Eq. (\ref{319}), the above equation takes the form 
\begin{eqnarray}
g_{\mu \nu }I^{\mu \nu }\left( q\right) &=&\frac{q^{2}}{2}\int \frac{d^{3}%
{\bf k}_{{\bf 1}}d^{3}{\bf k}_{{\bf 2}}}{E_{1}E_{2}}\delta ^{4}\left(
k_{1}+k_{2}-q\right)  \nonumber \\
&\equiv &\frac{1}{2}q^{2}I\left( q^{2}\right) .  \label{321}
\end{eqnarray}
We see from its defination that the integral $I\left( q^{2}\right) $ is an
invariant, so that it can be evaluated in any coordinate system. For our
convienence we shall choose the centre-of-mass system of two electrons. In
this system 
\[
{\bf k}_{{\bf 1}}=-{\bf k}_{{\bf 2}} 
\]
And for the massless electron 
\begin{eqnarray*}
E_{1} &=&\sqrt{\left| {\bf k}_{{\bf 1}}\right| ^{2}+m_{e}^{2}} \\
&=&\left| {\bf k}_{{\bf 1}}\right|
\end{eqnarray*}
also 
\[
E_{2}=\left| {\bf k}_{{\bf 2}}\right| . 
\]
So from above two results, we can write 
\begin{equation}
E_{1}=E_{2}=E.  \label{322}
\end{equation}
Hence by removing the integration on $k_{2}$, we get 
\begin{eqnarray}
I\left( q^{2}\right) &=&\int \frac{d^{3}{\bf k}_{{\bf 1}}}{E^{2}}\delta
\left( 2E-q_{0}\right)  \nonumber \\
&=&\int \frac{E^{2}dE}{E^{2}}\delta \left( 2E-q_{0}\right) d\Omega  \nonumber
\\
&=&2\pi  \label{323}
\end{eqnarray}
and from Eq. (\ref{321}) 
\begin{equation}
g_{\mu \nu }I^{\mu \nu }\left( q\right) =\pi q^{2}.  \label{324}
\end{equation}
Similarly, calculating $q_{\mu }q_{\nu }I^{\mu \nu }\left( q\right) $, one
gets 
\begin{eqnarray}
q_{\mu }q_{\nu }I^{\mu \nu }\left( q\right) &=&\int d^{3}{\bf k}_{{\bf 1}%
}d^{3}{\bf k}_{{\bf 2}}\frac{q_{\mu }q_{\nu }k_{1}^{\mu }k_{2}^{\nu }}{%
E_{1}E_{2}}\delta ^{4}\left( k_{1}+k_{2}-q\right)  \nonumber \\
&=&\int d^{3}{\bf k}_{{\bf 1}}d^{3}{\bf k}_{{\bf 2}}\frac{\left( q\cdot
k_{1}\right) \left( q\cdot k_{2}\right) }{E_{1}E_{2}}\delta ^{4}\left(
k_{1}+k_{2}-q\right) .  \label{325}
\end{eqnarray}
This is obtained by using Eq. (\ref{315}). Also using Eq. (\ref{319}), we
got 
\begin{eqnarray}
q_{\mu }q_{\nu }I^{\mu \nu }\left( q\right) &=&\frac{\left( q^{2}\right) ^{2}%
}{4}I\left( q^{2}\right)  \nonumber  \label{26} \\
&=&\frac{1}{2}\pi \left( q^{2}\right) ^{2}.  \label{327}
\end{eqnarray}
Substituting Eqs. (\ref{324}) and (\ref{327}) in Eq. (\ref{315}), we get 
\begin{eqnarray}
\pi \left( q^{2}\right) &=&4A\left( q^{2}\right) +q^{2}B\left( q^{2}\right)
\label{328} \\
\frac{1}{2}\pi \left( q^{2}\right) ^{2} &=&q^{2}A\left( q^{2}\right) +\left(
q^{2}\right) ^{2}B\left( q^{2}\right) .  \label{329}
\end{eqnarray}
Solving (\ref{328}) and (\ref{329}) for the vaules of $A\left( q^{2}\right) $
and $B\left( q^{2}\right) $, we have 
\begin{eqnarray*}
A\left( q^{2}\right) &=&\frac{\pi \left( q^{2}\right) }{6} \\
B\left( q^{2}\right) &=&\frac{2\pi }{6}.
\end{eqnarray*}
Hence Eq. (\ref{317}) leads to 
\begin{equation}
I^{\mu \nu }\left( q\right) =\frac{\pi }{6}\left( g^{\mu \nu }q^{2}+2q^{\mu
}q^{\nu }\right) .  \label{330}
\end{equation}

Then the partial muon decay rate can be written as 
\begin{eqnarray}
d\Gamma &=&\frac{1}{\left( 2\pi \right) ^{9}}\frac{1}{2E_{\mu }}\frac{d^{3}%
{\bf k}_{{\bf 3}}}{E_{3}}\int \frac{d^{3}{\bf k}_{{\bf 1}}d^{3}{\bf k}_{{\bf %
2}}}{E_{1}E_{2}}\sum_{spins=\pm \frac{1}{2}}\left| A\right| ^{2}\left( 2\pi
\right) ^{4}\delta ^{4}\left( k_{1}+k_{2}-q\right)  \nonumber \\
&=&\left( \frac{\tilde{g}_{V_{\mu e}}g_{V_{ee}}}{M_{V}^{2}}\right) ^{2}\frac{%
1}{\left( 2\pi \right) ^{5}}\frac{1}{2E_{\mu }}\frac{d^{3}{\bf k}_{{\bf 3}}}{%
E_{3}}  \nonumber  \label{31} \\
&&\times 2\left[ p_{\mu }k_{3\nu }I^{\mu \nu }\left( q\right) +k_{3\mu
}p_{\nu }I^{\mu \nu }\left( q\right) \right] .  \label{331}
\end{eqnarray}
Using (\ref{330}), (\ref{331}) becomes 
\begin{eqnarray}
d\Gamma &=&\left( \frac{\tilde{g}_{V_{\mu e}}g_{V_{ee}}}{M_{V}^{2}}\right)
^{2}\frac{1}{\left( 2\pi \right) ^{5}}\frac{1}{2E_{\mu }}\frac{d^{3}{\bf k}_{%
{\bf 3}}}{E_{3}}  \nonumber \\
&&\times 2\left[ 
\begin{array}{c}
p_{\mu }k_{3\nu }\left( g^{\mu \nu }q^{2}+2q^{\mu }q^{\nu }\right) \\ 
+k_{3\mu }p_{\nu }\left( g^{\mu \nu }q^{2}+2q^{\mu }q^{\nu }\right)
\end{array}
\right]  \nonumber \\
&=&\left( \frac{\tilde{g}_{V_{\mu e}}g_{V_{ee}}}{M_{V}^{2}}\right) ^{2}\frac{%
1}{\left( 2\pi \right) ^{5}}\frac{1}{2E_{\mu }}\frac{d^{3}{\bf k}_{{\bf 3}}}{%
E_{3}}\frac{\pi }{6}  \nonumber  \label{32} \\
&&\times 4\left[ \left( p\cdot k_{3}\right) q^{2}+2\left( k_{3}\cdot
q\right) \left( p\cdot q\right) \right] .  \label{332}
\end{eqnarray}
This is obtained by using Eq. (\ref{315}).

Finally, we must integrate Eq. (\ref{332}) over all momenta $k_{3}$ of the
emitted electron. For a muon at rest, i.e. in the rest frame of muon, we
have 
\begin{eqnarray}
p &=&\left( E_{\mu },0\right) =\left( m_{\mu },0\right)  \nonumber \\
k_{3} &=&\left( E_{3},{\bf k}_{{\bf 3}}\right) .  \label{333}
\end{eqnarray}
Also 
\begin{eqnarray}
q_{0} &=&m_{\mu }-E_{3}  \nonumber \\
{\bf q} &=&-{\bf k}_{{\bf 3}}  \label{334}
\end{eqnarray}
Now, we will calculate the terms involving in Eq. (\ref{332}), i.e. 
\begin{eqnarray}
q^{2} &=&\left( p+k_{3}\right) ^{2}  \nonumber \\
&=&\left( m_{\mu }^{2}-2m_{\mu }E_{3}\right) ,  \nonumber \\
k_{3}^{2} &=&0  \nonumber \\
k_{3}.p &=&m_{\mu }E_{3}  \nonumber \\
k_{3}.q &=&m_{\mu }E_{3}  \nonumber \\
p.q &=&m_{\mu }\left( m_{\mu }-E_{3}\right)  \label{335}
\end{eqnarray}
By substuting these values Eq. (\ref{332}) takes the form 
\begin{eqnarray}
d\Gamma &=&\left( \frac{\tilde{g}_{V_{\mu e}}g_{V_{ee}}}{M_{V}^{2}}\right)
^{2}\frac{1}{\left( 2\pi \right) ^{5}}\frac{1}{2m_{\mu }}\frac{d^{3}{\bf k}_{%
{\bf 3}}}{E_{3}}\frac{\pi }{6}4  \nonumber \\
&&\times \left[ \left( m_{\mu }E_{3}\right) \left( m_{\mu }^{2}-2m_{\mu
}E_{3}\right) +2\left( m_{\mu }E_{3}\right) m_{\mu }\left( m_{\mu
}-E_{3}\right) \right]  \nonumber \\
&=&\left( \frac{\tilde{g}_{V_{\mu e}}g_{V_{ee}}}{M_{V}^{2}}\right) ^{2}\frac{%
1}{\left( 2\pi \right) ^{5}}\frac{2\pi }{6}E_{3}^{2}dE_{3}d\Omega \left[
3m_{\mu }^{2}-4m_{\mu }E_{3}\right] .  \label{336}
\end{eqnarray}
Integrating Eq. (\ref{336}) over all directions $\Omega $ of the emitted
electron and over its complete range of energies $0\leq E_{3}\leq \frac{1}{2}%
m_{\mu }$, we obtain the total decay rated 
\begin{eqnarray}
\Gamma &=&\left( \frac{\tilde{g}_{V_{\mu e}}g_{V_{ee}}}{M_{V}^{2}}\right)
^{2}\frac{4\pi }{\left( 2\pi \right) ^{5}}\frac{2\pi }{6}m_{\mu }\int_{0}^{%
\frac{1}{2}m_{\mu }}E_{3}^{2}\left[ 3m_{\mu }-4E_{3}\right] dE_{3}  \nonumber
\\
&=&\left( \frac{\tilde{g}_{V_{\mu e}}g_{V_{ee}}}{M_{V}^{2}}\right) ^{2}\frac{%
4\pi }{\left( 2\pi \right) ^{5}}\frac{2\pi }{6}\frac{m_{\mu }^{5}}{16} 
\nonumber \\
\Gamma \left( \mu \rightarrow 3e\right) &=&\frac{1}{2}\left( \frac{\tilde{g}%
_{V_{\mu e}}g_{V_{ee}}}{M_{V}^{2}}\right) ^{2}\frac{m_{\mu }^{5}}{192\pi ^{3}%
}  \label{337}
\end{eqnarray}

Comparing the above contribution to the $\mu \rightarrow 3e$ process to that
of the ordionary muon decay, $\mu \rightarrow e\nu \bar{\nu}$, which
proceeds via $W$ exchange and (almost) identical kinematics, gives the
relation 
\begin{equation}
\frac{\Gamma \left[ \mu \rightarrow 3e\right] _{V-exch.}}{\Gamma \left[ \mu
\rightarrow e\nu \bar{\nu}\right] }\approx \frac{\tilde{g}_{V_{\mu
e}}^{2}g_{V_{ee}}^{2}}{M_{V}^{4}}/\frac{g_{W}^{4}}{M_{W}^{4}}.  \label{338}
\end{equation}
This ratio is defined as the branching ratio. Therefore 
\begin{equation}
\left[ {\cal B}\left( \mu \rightarrow 3e\right) \right] _{V-exch.}\approx 
\frac{\tilde{g}_{V_{\mu e}}^{2}g_{V_{ee}}^{2}}{M_{V}^{4}}/\frac{g_{W}^{4}}{%
M_{W}^{4}}.  \label{339}
\end{equation}
Since 
\[
\Gamma \left( V\rightarrow e^{+}e^{-}\right) \sim g_{V_{ee}}^{2}M_{V}, 
\]
and 
\[
\Gamma \left( V\rightarrow \mu ^{+}e^{-}\right) \sim \tilde{g}_{V_{\mu
e}}^{2}M_{V}, 
\]
while 
\[
\Gamma \left( W\rightarrow e\nu \right) \sim g_{W}^{2}M_{W}, 
\]
we can rewrite the above expression as 
\begin{equation}
\left[ {\cal B}\left( \mu \rightarrow 3e\right) \right] _{V-exch.}\approx 
\frac{\Gamma \left( V\rightarrow e^{+}e^{-}\right) \Gamma \left(
V\rightarrow \mu ^{+}e^{-}\right) }{\Gamma ^{2}\left( W\rightarrow e\nu
\right) }\left( \frac{M_{W}}{M_{V}}\right) ^{6}.  \label{340}
\end{equation}
Using $\left[ BR\left( \mu \rightarrow 3e\right) \right] _{V-exch.}\leq
10^{-12}$ and other data pertaining to the $e^{+}e^{-}$ widths of the
various vector mesons $V_{i}$, we find a set of bounds for the two body LFV
branching ratios of these vector mesons, i.e. 
\begin{equation}
\Gamma \left( V\rightarrow \mu ^{+}e^{-}\right) \approx \frac{\left[
BR\left( \mu \rightarrow 3e\right) \right] _{V-exch.}\Gamma ^{2}\left(
W\rightarrow e\nu \right) }{\Gamma \left( V\rightarrow e^{+}e^{-}\right) }%
\left( \frac{M_{V}}{M_{W}}\right) ^{6}.  \label{341}
\end{equation}
These bounds are calculated by dividing the above decay width by the full
width of the vector mesons. Using the numerical values these bounds become 
\begin{eqnarray}
{\cal B}\left( Z\rightarrow \mu e\right) &\leq &5\times 10^{-13}  \label{342}
\\
{\cal B}\left( J/\psi \rightarrow \mu e\right) &\leq &4\times 10^{-13}
\label{343} \\
{\cal B}\left( \Upsilon \rightarrow \mu e\right) &\leq &2\times 10^{-9}
\label{344} \\
{\cal B}\left( \Phi \rightarrow \mu e\right) &\leq &4\times 10^{-17}
\label{345}
\end{eqnarray}

All the vector (or psudoscalars) used as intermediaries in deriving the
bounds in Eqs. (\ref{342}--\ref{345}) are not on-shell. Thus we must
entertain the possibilty that their contribution to the three-body decays
considerd are reduced. This could at least weaken the various strong bounds
obtained above. Now we will focus on the possible mechanism for the
reduction of such bounds.

\subsection{Dynamical suppression of the LFV bounds}

The source of suppression is connected to possible ``form factor'' effect
due to the dynamics which would, for example, reduce the contribution of
various $V_{i}$ states to $\mu \rightarrow 3e$ compared to the naive
expectations. However the effect of these form factors should be minimal or
controllable if the LFV is induced by physics at scales much higher then the
EW scale of the $Z$ mass. These effects of dynamics are beutifully discussed
by Illuna, Jack and Rienmann in their recent papers \cite{0001273,0010193}.
Following these two papers, first of all we will discuss the dynamical
suppression in the two body decay of $Z\rightarrow l_{1}^{\pm }l_{2}^{\mp }$
and then move to $\mu \rightarrow 3e$.

The rare process $\mu \rightarrow e\gamma $ is the classic example of the
lepton flavour violating process. The previous limit on its branching ratio
is \cite{bolton} 
\[
{\cal B}\left( \mu \rightarrow e\gamma \right) <4.9\times 10^{-11}. 
\]
This reaction has not been observed so far, and the best experimental upper
limit of its branching fraction is \cite{0010193} 
\[
{\cal B}\left( \mu \rightarrow e\gamma \right) <1.2\times 10^{-11} 
\]
i.e. it is improved by the factor of $4.1$.

At the $Z$ factory LEP, searches for similar processes, through the $Z$
boson, became possible: 
\[
Z\rightarrow e\mu .
\]
The best experimental limit on its branching ratio is ($95\%$ C.L.) 
\[
{\cal B}\left( Z\rightarrow e^{\pm }\mu ^{\mp }\right) <1.7\times 10^{-6}.
\]
Let's discuss this reaction in detail.

The most general matrix element for the interaction of an on-shell vector
boson with a fermionic current, as shown in fig. \ref{tree3}, may be
described by four dimensionless form factors. At one loop level it is
convinent to parameterize the amplitude as 
\begin{equation}
A=-\frac{ig\alpha _{W}}{4\pi }\varepsilon ^{\rho }\bar{u}_{e}\left(
p_{2}\right) \Gamma _{\rho }v_{\mu }\left( p_{1}\right) ,  \label{3101}
\end{equation}
with $\alpha _{W}=\frac{g^{2}}{\left( 4\pi \right) }$, $\varepsilon $ being
the boson polarization vector and 
\begin{equation}
\Gamma _{\rho }=\gamma _{\rho }\left( f_{V}-f_{A}\gamma _{5}\right) +\frac{%
q^{\nu }}{M_{W}}\left( if_{M}+f_{E}\gamma _{5}\right) \sigma _{\rho \nu },
\label{3102}
\end{equation}
where 
\[
\sigma _{\rho \nu }=\frac{i}{2}\left( \gamma _{\rho }\gamma _{\nu }-\gamma
_{\nu }\gamma _{\rho }\right) 
\]
In Eq. (\ref{3102}), the form factors $f_{V}$ and $f_{A}$ stands for vector
and axial-vector couplings and $f_{M}$ and $f_{E}$ for magnetic and electric
dipole moments/transitions of equal/unlike final fermions. The form factors
depend on the momentum transfer squared $Q^{2}=\left( p_{2}+p_{1}\right) ^{2}
$. In principle all the four form factors must be non-zero. In order to find
the decay width we have to find $|A|^{2}$ which is 
\begin{equation}
|A|^{2}=AA^{\dagger }.  \label{3103}
\end{equation}
So 
\begin{eqnarray}
|A|^{2} &=&\left. \left( \frac{g\alpha _{W}}{4\pi }\right) ^{2}\right[
8\left( p_{2}\cdot p_{1}\right) \left\{ \left| f_{V}\right| ^{2}+\left|
f_{A}\right| ^{2}\right\}   \nonumber  \label{08} \\
&&+\left. \frac{\left\{ \left| f_{E}\right| ^{2}+\left| f_{M}\right|
^{2}\right\} }{M_{W}^{2}}\left\{ 16\left( p_{2}\cdot q\right) \left(
p_{1}\cdot q\right) -4\left( q\cdot q\right) \left( p_{2}\cdot p_{1}\right)
\right\} \right] .  \nonumber \\
&&  \label{3108}
\end{eqnarray}
Using the spin averages, above result becomes 
\begin{eqnarray}
\sum_{spins}|A|^{2} &=&\left. \left( \frac{g\alpha _{W}}{4\pi }\right) ^{2}%
\frac{1}{12}\right[ 8\left( p_{2}\cdot p_{1}\right) \left\{ \left|
f_{V}\right| ^{2}+\left| f_{A}\right| ^{2}\right\}   \nonumber \\
&&+\left. \frac{\left\{ \left| f_{E}\right| ^{2}+\left| f_{M}\right|
^{2}\right\} }{M_{W}^{2}}\left\{ 16\left( p_{2}\cdot q\right) \left(
p_{1}\cdot q\right) -4\left( q\cdot q\right) \left( p_{2}\cdot p_{1}\right)
\right\} \right] .  \nonumber \\
&&  \label{3109}
\end{eqnarray}

The decay width can be calculated as 
\begin{eqnarray}
\Gamma &=&\frac{1}{2M_{Z}}\int \int \frac{d^{3}{\bf p}_{1}}{\left( 2\pi
\right) ^{3}E_{1}}\frac{d^{3}{\bf p}_{2}}{\left( 2\pi \right) ^{3}E_{2}}%
\sum_{spins}|A|^{2}\left( 2\pi \right) ^{4}\delta ^{4}\left(
q-p_{1}-p_{2}\right)  \nonumber \\
&=&\frac{\alpha _{W}^{3}}{24\pi ^{2}}M_{Z}\left[ \left| f_{V}\right|
^{2}+\left| f_{A}\right| ^{2}+\frac{1}{2c_{W}^{2}}\left( \left| f_{E}\right|
^{2}+\left| f_{M}\right| ^{2}\right) \right] .  \label{3110}
\end{eqnarray}
The branching ratio can be obtained by dividing the decay width obtained in
Eq. (\ref{3110}) by the total width of $Z$ boson. 
\begin{equation}
{\cal B}\left[ Z\rightarrow \mu e\right] =\frac{\alpha _{W}^{3}}{24\pi ^{2}}%
\frac{M_{Z}}{\Gamma _{Z}}\left[ \left| f_{V}\right| ^{2}+\left| f_{A}\right|
^{2}+\frac{1}{2c_{W}^{2}}\left( \left| f_{E}\right| ^{2}+\left| f_{M}\right|
^{2}\right) \right] .  \label{3111}
\end{equation}
where 
\begin{eqnarray*}
\Gamma _{Z} &\approx &\frac{\alpha _{W}}{c_{W}^{2}}M_{Z} \\
c_{W} &=&\frac{M_{W}}{M_{Z}}.
\end{eqnarray*}
These form factors are model dependent. It is easily seen that the branching
fraction can be approximated as 
\begin{equation}
{\cal B}\left( Z\rightarrow \mu e\right) \sim \left( \frac{\alpha _{W}}{\pi }%
\right) ^{2}\sim 10^{-6}  \label{3112}
\end{equation}
This is in agreement with its experimental value. Now we shall discuss these
suppression in $\mu \rightarrow 3e$ decay.

CASE-1:

First discuss the vector-axial-vector form factor contribution at $V_i\mu e$
vertex to $\mu \rightarrow 3e$, i.e. instead of the $\gamma ^\alpha $ we
have $\gamma ^\alpha \left( f_V+f_A\gamma _5\right) $. Therefore the
amplitude becomes 
\begin{eqnarray}
A\left( \mu \rightarrow 3e\right) &=&\bar{u}_e(k_3)\gamma ^\alpha \left(
f_V+f_A\gamma _5\right) u_\mu (p)\bar{u}_e(k_2)\gamma _\alpha v_e(k_1)\frac{%
\tilde{g}_{V_{\mu e}}g_{V_{ee}}}{M_V^2-s}  \nonumber  \label{13} \\
&=&\bar{u}_e(k_3)\gamma ^\alpha \left( f_V+f_A\gamma _5\right) u_\mu (p)\bar{%
u}_e(k_2)\gamma _\alpha v_e(k_1)\frac{\tilde{g}_{V_{\mu e}}g_{V_{ee}}}{M_V^2}
\nonumber \\
&&  \label{3113}
\end{eqnarray}
Using the same procedure as for the previous cases $\left| A\right| ^2$, we
get 
\begin{equation}
\left| A\right| ^2=32\left[ \left| f_V\right| ^2+\left| f_A\right| ^2\right]
\left[ \left( k_1.p\right) \left( k_2.k_3\right) +\left( k_1.k_3\right)
\left( k_2.p\right) \right] \left( \frac{\tilde{g}_{V_{\mu e}}g_{V_{ee}}}{%
M_V^2}\right) ^2  \label{3114}
\end{equation}
Using the spin averages, the above expression becomes 
\begin{equation}
\sum_{spins=\pm \frac 12}\left| A\right| ^2=2\left[ \left| f_V\right|
^2+\left| f_A\right| ^2\right] \left[ \left( k_1.p\right) \left(
k_2.k_3\right) +\left( k_1.k_3\right) \left( k_2.p\right) \right] \left( 
\frac{\tilde{g}_{V_{\mu e}}g_{V_{ee}}}{M_V^2}\right) ^2.  \label{3115}
\end{equation}
Solving the delta function using the technique described above, the decay
width becomes 
\begin{equation}
\Gamma \left( \mu \rightarrow 3e\right) =\frac 12\left[ \left| f_V\right|
^2+\left| f_A\right| ^2\right] \left( \frac{\tilde{g}_{V_{\mu e}}g_{V_{ee}}}{%
M_V^2}\right) ^2\frac{m_\mu ^5}{192\pi ^3}.  \label{3116}
\end{equation}
The corresponding branching ratio becomes 
\begin{equation}
{\cal B}\left( \mu \rightarrow 3e\right) \approx \left[ \left| f_V\right|
^2+\left| f_A\right| ^2\right] \frac{\tilde{g}_{V_{\mu e}}^2g_{V_{ee}}^2}{%
M_V^4}/\frac{g_W^4}{M_W^4}.  \label{3117}
\end{equation}

CASE-2:

Now consider the electric and magnetic form factor contribution to the $\mu
\rightarrow 3e$ at $V_{i}\mu e$ vertex. The amplitude can be written as 
\begin{equation}
A=\bar{u}_{e}(k_{3})\frac{q^{\nu }}{M_{W}}\left( if_{M}+f_{E}\gamma
_{5}\right) \sigma _{\alpha \nu }u_{\mu }(p)\bar{u}_{e}(k_{2})\gamma
^{\alpha }v_{e}(k_{1})\frac{\tilde{g}_{V_{\mu e}}g_{V_{ee}}}{M_{V}^{2}}.
\label{3119}
\end{equation}
Using the same procedure as before the decay width for this reaction becomes 
\begin{equation}
\Gamma \left( \mu \rightarrow 3e\right) =\frac{1}{2}\frac{1}{80}\left( \frac{%
\tilde{g}_{V_{\mu e}}g_{V_{ee}}}{M_{V}^{2}}\right) ^{2}\left[ \left|
f_{M}\right| ^{2}+\left| f_{E}\right| ^{2}\right] \frac{m_{\mu }^{2}}{%
M_{W}^{2}}\frac{m_{\mu }^{5}}{192\pi ^{3}}  \label{3122}
\end{equation}
The branching ratio for this case becomes 
\begin{equation}
{\cal B}\left( \mu \rightarrow 3e\right) \approx \frac{1}{80}\left[ \left|
f_{M}\right| ^{2}+\left| f_{E}\right| ^{2}\right] \frac{m_{\mu }^{2}}{%
M_{W}^{2}}\left( \frac{\tilde{g}_{V_{\mu e}}^{2}g_{V_{ee}}^{2}}{M_{V}^{4}}/%
\frac{g_{W}^{4}}{M_{W}^{4}}\right) .  \label{3123}
\end{equation}
Hence, the branching ratios of the three body decays are suppressed by a
factor $\frac{m_{\mu }^{2}}{M_{W}^{2}}.$

\newpage 

%
%
%
%
\begin{figure}
\label{tree2}
\caption{$\mu \rightarrow 3 e$}
\end{figure}

\begin{figure}
\label{tree3}
\caption{$Z \rightarrow \mu e$}
\end{figure}

%

\chapter{Lepton Flavour Violation in LSND and Seesaw model of neutrino masses
}

\section{Introduction}

The question of whether or not neutrinos have a nonzero mass has remained
one of the most tantalizing issue in the present day physics. In the
Standard Model of electroweak theory neutrinos are considerd to be massless.
But there is no compelling theoratical reason behind this assumption \cite
{PRD55-2931}. Hints that the neutrinos are massive particles comes from the
observation of three anomalous efects,

\begin{enumerate}
\item  the LSND (Liquid Scentilator Neutrino Detector) excess \cite
{nucl-ex/9605001, nucl-ex/9706006,hep-ph/9809524},

\item  the atmospheric anomaly \cite{Phys.Lett B 335 237(1994),
hep-ex/0009001,Phys. Lett. B 449 137(1999)} and

\item  the solar neutrino deficit \cite{Astrophys.J.496,Phys. Rev. Lett.
77,Phys. Lett. B447,astro-ph/9907113}.
\end{enumerate}

In particular, the atmospheric results are the most convincing ones. All the
three effects can be naturally explained in terms of neutrino flavour
oscillations, which will occur when neutrino propagate through space, if
there masses are non-degnerate and the weak and mass eigenstates are mixed.

\subsection{The neutrino flavour oscillation}

However, in order to explain all three experimentally observed effects in
terms of neutrino flavour oscillation, one is forced to invoke additional
sterile neutrino states \cite{hep-ph/9910336 } to accommodate the very
different frequencies of oscillations, given by three different mass squared 
$\Delta m^{2}$'s, indicated by three different effects. The existance of
such neutrino is currently unresolved problem and clearly demonstrates that
the neutrino sector is not fully understood. So due to the unappealing
theoretical feature of a light sterile neutrino, it is interesting to look
for alternatives that could explain the LSND excess with the known three
light neutrinos.

From a phenomenological point of view, we recall that the neutrino flavour
oscillation hypothesis predicts a well defined dependence of phenomena as a
function of neutrino energy, characterized by the so called $L/E$ behaviour,
where $L$ is distance between source and detector and $E$ is neutrino
energy. So far, no experiment has conclusively demonstrated such a $L/E$
dependence of the anomalous effect, with may be expectation of
SuperKamiokande data which favours the dependence $\alpha $ $LE^n$ where $%
n=-1$ \cite{Nov.1 2000}.

In such an unclear sitution, it is possible to envisage ``non--flavour
oscillation'' mechanism to explain part of neutrino data.

Aside from theoretical argument against sterile neutrino we argue that, from
a phenomenological point of view, the LSND effect is peculiar:\ it has a
small probability, measured to be $\left( 2.5\pm 0.6\pm 0.4\right) \times
10^{-3}$ \cite{Nov.1 2000}, in contrast to the solar and atmospheric
neutrino anomalies, which are large. Hence, LSND is a natural candidate for
an interpretation involving a different physics than in atmospheric and
solar neutrino flavour oscillation.

\section{LSND\ puzzle}

We recall that the LSND\ effect was first reported as an excess of $\bar{\nu}%
_{e}$'s in the $\bar{\nu}_{\mu }$ flux from the $\mu ^{+}$ decay at rest
(DAR) process. The neutrino beam is obtained with $800$ kinetic energy
protons hitting a series of targets, producing secondary pions. Most of $\pi
^{+}$ come to rest and decay through the sequence $\pi ^{+}\rightarrow \mu
^{+}\nu _{\mu }$; followed by $\mu ^{+}\rightarrow e^{+}\nu _{e}\bar{\nu}%
_{\mu }$, supplying the experiment with the $\bar{\nu}_{\mu }$'s with a
maximum energy of $52.8$ $MeV$. The intrisnic contamination of $\bar{\nu}%
_{e} $'s coming from the symmetrical decay chain starting with $\pi ^{-}$ is
estimated to be small since most negatively charged mesons are captured
before they decay.

The excess of $\bar{\nu}_e$'s, explained in terms of neutrino flavour
transitions of the type $\bar{\nu}_\mu \rightarrow \bar{\nu}_e$, occur via
the reactions: 
\begin{eqnarray*}
&&\left. \mu ^{+}\rightarrow e^{+}\nu _e\bar{\nu}_\mu \right. \\
&&\bar{\nu}_\mu \stackrel{vacuum}{\longrightarrow }\bar{\nu}_e \\
&&\left. \bar{\nu}_ep\rightarrow e^{+}n\right.
\end{eqnarray*}
i.e. anti-neutrinos are produced by the $\mu ^{+}\rightarrow e^{+}\nu _e\bar{%
\nu}_\mu $ and detected by $\bar{\nu}_ep\rightarrow e^{+}n$.

There is another evidence in favour of the neutrino flavour oscillation
which was reported in the Decay In Flight (DIF). But due to its lower
statistical significance, we concentrate on the hint from stopped muon, and
ignore the DIF result.

The LSND claim is contradicted by the latest KARMEN2 results \cite
{hep-ex/0008002}, however the experimental sensitivity is marginal to
conclusively exclude or confirm completely the LSND\ excess. A new
experiment, MiniBOONE \cite{hep-ex/0009056} would confornt the flavour
oscillation hypothesis with a very high statistical accuracy. A negative
result from MiniBOONE\ experiment would indicate that the neutrino flavour
oscillation is not the correct hypothesis to explain the excess seen in
LSND. It would however not contradict other possible non-flavour-oscillation
interpretation of the effect. In particular, LFV decays would play a role in
the interpretation of LSND excess. A neutrino factory is an ideal mechaine
to probe such anomalous decays of muon.

Th aim of this work is to investigate another approch. We assume that the
three neutrinos interact through Lepton Flavour Violation interactions,
which are forbidden in the Standard Model. This is an attractive
possibility, because various extensions of the Standard Model which predict
the neutrino masses also give rise to such new interactions. These
interactions can affect the LSND excess. The motivation for the LFV is that
the branching probability for LSND\ is too small ($\left( 2.5\pm 0.6\pm
0.4\right) \times 10^{-3})$) compared to atmospheric and solar anomalies. We
analyze the consequence of small Lepton Flavour Violating interactions to
explain the LSND excess, and check whether this scenario will be feasible.

\subsection{Basic consideration and calculation}

Let us assume that a vector boson $V_{i}$ (here $V_{i}$ could be either a
fundamental state, such as the $Z^{0}$ or the quark- antiquark bound state
such as the $\phi ,J/\psi ,$ or $\Upsilon $) couples to $\mu ^{\pm }e^{\pm }$%
. If it couples also to $e^{+}e^{-}$, as all the states above do, then by
unitarity its exchange contributes also to $\mu \rightarrow 3e$. Let us
define the effctive couplins between vector boson $V_{i}$ and $\mu ^{\pm
}e^{\mp }$ as $\tilde{g}_{V_{\mu e}}$, and the corresponding effective
Lagrangian can be written as 
\begin{equation}
{\cal L}_{eff}=\tilde{g}_{V_{\mu e}}\bar{\mu}\gamma ^{\alpha }eV^{\alpha
}+H.c.  \label{401}
\end{equation}
Here, the reaction under consideration is 
\[
\mu ^{+}\rightarrow e^{+}\nu _{l}\bar{\nu}_{l},
\]
therefore amoung these vector bosons only the $Z^{0}$ will contribute. $l$
is any of the three known leptons, i.e. $e$, $\mu $, or $\tau $. The
coupling at the LFV vertex remains the same, but at the neutrino vertex it
will be changed and is obtained by the Standard Model. We also assume that
the LFV decay proceds through a similar diagram as the Standard Model muon
decay, however with interchanged neutrino flavour as shown in fig. \ref
{tree4}. 

Using Feynman rules, the corresponding amplitude can be written as 
\begin{eqnarray}
A\left( \mu \rightarrow e\nu _{l}\bar{\nu}_{l}\right) &=&\frac{\tilde{g}%
_{Z_{\mu e}}g_{Z\nu \bar{\nu}}g_{\alpha \beta }}{M_{Z}^{2}-s}\left[ \bar{v}%
(p_{1})\gamma ^{\alpha }v(p_{2})\bar{u}(p_{4})\gamma ^{\beta }\left(
1-\gamma _{5}\right) v(p_{3})\right]  \nonumber   \\
&=&\frac{\tilde{g}_{Z_{\mu e}}g_{Z\nu \bar{\nu}}}{M_{Z}^{2}-s}\left[ \bar{v}%
(p_{1})\gamma ^{\alpha }v(p_{2})\bar{u}(p_{4})\gamma _{\alpha }\left(
1-\gamma _{5}\right) v(p_{3})\right]  \label{402}
\end{eqnarray}
Here $g_{V\nu \bar{\nu}}$ is the effective coupling of the $Z$- boson to $%
\nu \bar{\nu}$. Also for low energy, we have to neglect $s$ in comparison to
the $M_{Z}$. Therefore Eq. (\ref{402}) takes the form 
\begin{equation}
A\left( \mu \rightarrow e\nu _{l}\bar{\nu}_{l}\right) =\frac{\tilde{g}%
_{Z_{\mu e}}g_{Z\nu \bar{\nu}}}{M_{Z}^{2}}\left[ \bar{v}(p_{1})\gamma
^{\alpha }v(p_{2})\bar{u}(p_{4})\gamma _{\alpha }\left( 1-\gamma _{5}\right)
v(p_{3})\right]  \label{403}
\end{equation}
Now as usual we have to calculate $\left| A\right| ^{2}$, which is 
\begin{eqnarray}
\sum_{spins=\pm \frac{1}{2}}\left| A\right| ^{2} &=&\frac{1}{16}\left( \frac{%
\tilde{g}_{Z_{\mu e}}g_{Z\nu \bar{\nu}}}{M_{Z}^{2}}\right) ^{2}Tr\left\{ \not%
{p}_{1}\gamma ^{\alpha }\not{p}_{2}\gamma ^{\beta }\right\}  \nonumber
 \\
&&\times Tr\left\{ \not{p}_{4}\gamma _{\alpha }\left( 1-\gamma _{5}\right) 
\not{p}_{3}\gamma _{\beta }\left( 1-\gamma _{5}\right) \right\} .
\label{406}
\end{eqnarray}
Solving these traces, the above Eq. becomes 
\begin{equation}
\sum_{spins=\pm \frac{1}{2}}\left| A\right| ^{2}=\left( \frac{\tilde{g}%
_{Z_{\mu e}}g_{Z\nu \bar{\nu}}}{M_{Z}^{2}}\right) ^{2}4\left[ \left(
p_{1}p_{4}\right) \left( p_{2}p_{3}\right) +\left( p_{1}p_{3}\right) \left(
p_{2}p_{4}\right) \right] .  \label{407}
\end{equation}
Then the partial decay rate for the muon can be written as 
\begin{equation}
d\Gamma =\frac{1}{\left( 2\pi \right) ^{9}}\frac{1}{2E_{\mu }}\frac{d^{3}%
{\bf k}_{3}}{E_{3}}\int \frac{d^{3}{\bf k}_{1}d^{3}{\bf k}_{2}}{E_{1}E_{2}}%
\sum_{spins=\pm \frac{1}{2}}\left| A\right| ^{2}\left( 2\pi \right)
^{4}\delta ^{4}\left( p_{1}+p_{2}-q\right) ,  \label{408}
\end{equation}
where 
\[
q=p_{3}+p_{4.} 
\]
Using the same technique as we have used in the previous chapter, we got the
total decay width as follows 
\begin{equation}
\Gamma \left( \mu \rightarrow e\nu _{l}\bar{\nu}_{l}\right) =\left( \frac{%
\tilde{g}_{Z_{\mu e}}g_{Z\nu \bar{\nu}}}{M_{Z}^{2}}\right) ^{2}\frac{m_{\mu
}^{5}}{192\pi ^{3}}.  \label{409}
\end{equation}

Comparing the above contribution to the $\mu \rightarrow e\nu _l\bar{\nu}_l$
process to that of the ordionary muon decay, $\mu \rightarrow e\nu \bar{\nu}$%
, which proceeds via $W$ exchange and (almost) identical kinematics, gives
the relation 
\begin{eqnarray}
\frac{\Gamma \left( \mu \rightarrow e\nu _l\bar{\nu}_l\right) _{Z-exch.}}{%
\Gamma (\mu \rightarrow e\nu \bar{\nu})} &=&\left[ \frac{\Gamma \left( \mu
\rightarrow e\nu _l\bar{\nu}_l\right) _{Z-exch.}}{\Gamma \left[ \mu
\rightarrow 3e\right] _{V-exch.}}\right] \times \left[ \frac{\Gamma \left(
\mu \rightarrow 3e\right) _{V-exch.}}{\Gamma (\mu \rightarrow e\nu \bar{\nu})%
}\right]  \nonumber \\
{\cal B}\left( \mu \rightarrow e\nu _l\bar{\nu}_l\right) _{Z-exch.}
&=&\left[ \frac{\Gamma \left( \mu \rightarrow e\nu _l\bar{\nu}_l\right)
_{Z-exch.}}{\Gamma \left[ \mu \rightarrow 3e\right] _{V-exch.}}\right]
\times {\cal B}\left( \mu \rightarrow 3e\right) _{V-exch.}  \label{410}
\end{eqnarray}
Using the value of the $\Gamma \left[ \mu \rightarrow 3e\right] _{Z-exch.}$,
Eq. (\ref{410}) becomes 
\begin{equation}
{\cal B}\left( \mu \rightarrow e\nu _l\bar{\nu}_l\right) _{Z-exch.}\approx
\left( \frac{g_{Z\nu \bar{\nu}}}{g_{Zee}}\right) ^2\times {\cal B}\left( \mu
\rightarrow 3e\right) _{V-exch.}.  \label{411}
\end{equation}
We know that 
\[
\left[ {\cal B}\left( \mu \rightarrow 3e\right) \right] _{V-exch.}\leq
10^{-12}. 
\]
Since 
\[
\Gamma \left( Z\rightarrow e^{+}e^{-}\right) \sim g_{Z_{ee}}^2M_Z 
\]
and 
\[
\Gamma \left( Z\rightarrow \nu _l\bar{\nu}_l\right) \sim g_{Z\nu \bar{\nu}%
}^2M_Z. 
\]
So Eq. (\ref{410}) becomes 
\begin{eqnarray}
{\cal B}\left( \mu \rightarrow e\nu _l\bar{\nu}_l\right) _{Z-exch.} &\leq &%
\frac{\Gamma \left( Z\rightarrow \nu _l\bar{\nu}_l\right) }{\Gamma \left(
Z\rightarrow e^{+}e^{-}\right) }\times 10^{-12}  \nonumber  \label{12} \\
&\leq &2\times 10^{-12}.  \label{412}
\end{eqnarray}
As it is seen from the above equation the branching ratio for the this
lepton flavour violating decay is the same as that of the $\mu \rightarrow
3e $. This is because of the fact that the reaction kinematic is same , and
also in both the cases the external particles are assumed to be the
massless. The factor of $2$ is due to the fact that the couplings between
electron-positron is different then the two neutrinos.

It is already mentioned that, in order to explain the LSND result, the
effective New Physics coupling should satisfy 
\begin{equation}
r=\left| \frac{G_N^\nu }{G_F}\right| ^2=\left( 2.5\pm 0.6\pm 0.4\right)
\times 10^{-3}.  \label{413}
\end{equation}
We define $G_N^\nu $ to be the effective coupling of the anomalous muon
decay. Thus, at the $90\%$ C.L. we need \cite{hep-ph/9809524} 
\begin{equation}
r>1.6\times 10^{-3}.  \label{414}
\end{equation}
Now let's calculate this for the reaction discussed here. As we know that 
\begin{equation}
\frac{G_F}{\sqrt{2}}=\frac{g_W^2}{M_W^2}.  \label{415}
\end{equation}
Thus following the Eq. (\ref{414}), we can write the effective coupling as
follows 
\begin{equation}
\frac{G_F^{eff}}{\sqrt{2}}=\frac{\tilde{g}_{Z_{\mu e}}g_{Z\nu \bar{\nu}}}{%
M_Z^2}.  \label{416}
\end{equation}
Now the ratio of the two couplings become 
\begin{eqnarray}
\frac{G_F^{eff}}{G_F}=\frac{\tilde{g}_{Z_{\mu e}}g_{Z\nu \bar{\nu}}}{M_Z^2}%
\frac 1{G_F}  \label{417}
\end{eqnarray}
So $r$ becomes 
\begin{eqnarray}
r &=&\left| \frac{G_F^{eff}}{G_F}\right| ^2  \nonumber \\
&=&\frac{\Gamma \left( \mu \rightarrow 3e\right) _{Z-exch.}}{\Gamma (\mu
\rightarrow e\nu \bar{\nu})}\frac 12\frac{g_{Z\nu \bar{\nu}}^2}{%
g_{Ze^{+}e^{-}}^2}  \label{418}
\end{eqnarray}
As and using the value of the $g_{Z\nu \bar{\nu}}^2$ and $g_{Ze^{+}e^{-}}^2$
from the Standard Model, we get 
\begin{equation}
r\sim 10^{-12}.  \label{419}
\end{equation}
Thus, comparing with Eq. (\ref{414}) which requires $r>1.6\times 10^{-3}$,
we learn that the LFV anomalous muon decays $\mu \rightarrow e\nu _l\bar{\nu}%
_l$ cannot significantly contribute to the LSND DAR result, at least when we
use $\tilde{g}_{Z_{\mu ^{+}e^{-}}}$ as constrained by $\mu \rightarrow 3e$
in a model independent way.

Thus the excess found by the LSND is not due to these Lepton Flavour
Violating decays. So we cannot say that whether this excess is due to
neutrino flavour oscillations or due to LFV decays which might occur through
some exotic mechanism leading to $\mu ^{+}\rightarrow e^{+}\bar{\nu}_{e}\bar{%
\nu}_{l}$. Hence, we can say that it is still unresolved puzzle of the
neutrino physics.

\section{Seesaw model of neutrino masses}

In the minimal standard gauge model of quarks and leptons, each of three
known neutrinos $\left( \nu _e\text{, }\nu _\mu \text{, }\nu _\tau \right) $
appears only as a member of left-handed $SU(2)$ lepton doublet 
\begin{equation}
\psi _i=\left( \nu _i,l_i\right) _L,  \label{4101}
\end{equation}
and the Higgs sector contains only one scalar doublet 
\begin{equation}
\Phi =\left( \phi ^{+},\phi ^0\right) .  \label{4102}
\end{equation}
As a result, neutrinos are massless in this model. Experimentally there is
now a host of evidence for neutrino oscillations, and that is most naturally
explained if neutrinos are massive and mix with each other. Theoretically
there is no compelling reason for massless neutrinos, and any extension
beyond the minimal Standard Model often allows them to massive. There exists
already a vast literature on specific models of neutrino masses and mixing 
\cite{Phys. Rev. Lett Vol 81(1171)1998}.

Here we will make the following simple observations. In the minimal Standard
Model of particle interactions, the massless neutrinos acquire naturally
small Majorana masses through the effective dimension-five operator 
\begin{equation}
\frac{1}{\Lambda }\left( \nu _{i}\phi ^{0}-l_{i}\phi ^{+}\right) \left( \nu
_{j}\phi ^{0}-l_{j}\phi ^{+}\right) ,  \label{4103}
\end{equation}
where $\Lambda $ is a large effective mass scale, and $\Phi =\left( \phi
^{+},\phi ^{0}\right) $ is the usual Higgs doublet with the non zero
expectation value, $\left\langle \phi ^{0}\right\rangle =v$. All models of
neutrino mass and mixing (which have the same light particle content as at
the minimal Standard Model) can be explained by the operator 
\begin{equation}
\Lambda ^{-1}\phi ^{0}\phi ^{0}\nu _{i}\nu _{j}.  \label{4104}
\end{equation}
Different models are merely different reasons for this operator. The
intermediate heavy particle in case of the operator defined in Eq. (\ref
{4103}) is clearly a fermion singlet. Let's call it $N_{i}$ and let its mass
be $m_{N}$ and its coupling to $\nu _{i}$ be $f_{i}$. Also we can identify $%
f_{i}v$ or simply $fv$ as a Dirac mass $m_{D}$ linking $\nu _{i}$ to $N_{i}$
and the neutrino mass matrix is introduced so that 
\begin{equation}
m_{\nu }=\frac{m_{D}^{2}}{m_{N}},  \label{4105}
\end{equation}
so that $\Lambda =m_{N}/f^{2}$ in Eq. (\ref{4103}). This is, of course, just
the well known canonical seesaw mechanism, with $N_{i}$ identified as the
right-handed neutrino with a large Majorana mass. Given that $m_{\nu }$ is
at most of order $1eV$ and $f$ should be too small, the usual thinking is
that $m_{N}$ has to be very large, i.e., $m_{N}\gg v$. As such, this famous
mechanism must be accepted on faith, because there cannot be any direct
experimental test of its validity.

Consider now the possibility that there is no new physics beyond the $TeV$
energy scale. This is an intriguing idea proposed recently in theories of
large extra dimensions \cite{N. Arkani Hamid PLB 429 263 (1998)}. Instead of
using an ingradient supplied by the large extra dimensions, a model has
recently been proposed \cite{PRL. 86 2502 (2001)} to show how Eq. (\ref{4105}%
) may be realized naturally with $m_{N}$ of order $1$ $TeV$ in a simple
extension of the Standard Model. This means that $m_{D}$ should be small
i.e., $m_{D}\ll 10^{2}GeV$. If it comes from zeorth component of $\Phi $
i.e., $\phi ^{0}$ as in the Standard Model, that would not be natural; but
as shown below, it will come instead from another with naturally small
vacuum expectation value. This new realization of the seesaw mechanism will
allow direct experimental test of its validity, as discussed below.

Consider the minimal Standard Model with three lepton families: 
\begin{equation}
\binom{\nu _i}{l_i}_L\sim \left( 1,2,-1/2\right) ,\qquad l_{iR}\sim \left(
1,1,-1\right) ,  \label{4106}
\end{equation}
where their transformations under the standard $SU(3)_C\times SU(2)_L\times
U(1)_Y$ gauge group are indicated above. Now add three neutral fermion
singlets 
\begin{equation}
N_{iR}\sim \left( 1,1,0\right) ,  \label{4107}
\end{equation}
but instead of assigning them the lepton number $L=1$, so that they can pair
up with the lepton doublet through the interaction $\bar{N}_R\left( \nu
_L\phi ^0-l_L\phi ^{+}\right) $, $L=0$ is assigned to forbid this Yukawa
term. To complete the model, a new scalar doublet, called leptoquark \cite
{PRL Vol.79 (22) 4318(1997)} 
\begin{equation}
\binom{\eta ^{+}}{\eta ^0}\sim \left( 1,2,1/2\right)  \label{4108}
\end{equation}
is introduced with lepton number $L=-1$. Hence the terms 
\begin{equation}
\frac 12M_iN_{iR}^2+f_{ij}\bar{N}_{iR}\left( \nu _{jL}\eta ^0-l_{jL}\eta
^{+}\right) +H.c.  \label{4109}
\end{equation}
appear in the Lagrangian. The effective operator of Eq. (\ref{4103}) for
neutrino mass is then replaced by one with $\eta $ instead of $\phi $, and,
if $\left\langle \eta ^0\right\rangle =u$ is naturally small, the
corresponding scale $\Lambda $ will not have to be so large and $M_i$ of Eq.
(\ref{4109}) may indeed be of order $1$ $TeV$.

The Higgs potential of this model is given by 
\begin{eqnarray}
V &=&m_1^2\Phi ^{\dagger }\Phi +m_2^2\eta ^{\dagger }\eta +\frac 12\lambda
_1\left( \Phi ^{\dagger }\Phi \right) ^2+\frac 12\lambda _2\left( \eta
^{\dagger }\eta \right) ^2  \label{4110} \\
&&+\lambda _3\left( \Phi ^{\dagger }\Phi \right) \left( \eta ^{\dagger }\eta
\right) +\lambda _4\left( \Phi ^{\dagger }\eta \right) \left( \eta ^{\dagger
}\Phi \right) +\mu _{12}^2\left( \Phi ^{\dagger }\eta +\eta ^{\dagger }\Phi
\right) ,  \nonumber
\end{eqnarray}
where the $\mu _{12}^2$ term breaks $L$ explicitly but softly. Note that,
given the particle content of this model, the $\mu _{12}^2$ term is the only
possible soft term which also breaks $L$.

The equations of constraints for $\left\langle \phi ^0\right\rangle =v$ and $%
\left\langle \eta ^0\right\rangle =u$ are 
\begin{eqnarray}
v\left[ m_1^2+\lambda _1v^2+\left( \lambda _3+\lambda _4\right) u^2\right]
+\mu _{12}^2u &=&0  \label{4111} \\
u\left[ m_2^2+\lambda _2u^2+\left( \lambda _3+\lambda _4\right) v^2\right]
+\mu _{12}^2v &=&0.  \label{4112}
\end{eqnarray}
Consider the case 
\begin{equation}
m_1^2<0,\qquad m_2^2>0,\qquad \left| \mu _{12}^2\right| \ll m_2^2,
\label{4113}
\end{equation}
then 
\begin{equation}
v^2\simeq -\frac{m_1^2}{\lambda _1},\qquad u\simeq -\frac{\mu _{12}^2v}{%
m_2^2+\left( \lambda _3+\lambda _4\right) v^2}.  \label{4114}
\end{equation}
Hence it is very clear that $u$ may be very small compared to $v(=174$ $GeV)$%
. For example, if $m_2\sim 1TeV$, $\left| \mu _{12}^2\right| \sim 10GeV^2$,
then $u\sim 1MeV$. The relative smallness of $\left| \mu _{12}^2\right| $
may be attributed to the fact that it corrosponds to the explecit breaking
of the lepton number in $V$ of Eq. (\ref{4110}). The usual argument here is
that, if $\left| \mu _{12}^2\right| $ were zero, then as seen from Eq. (\ref
{4114}) the model's symmetry is increased, i.e., the lepton number would not
be broken. Hence the assumption that it is small compared to $\left|
m_1^2\right| $ or $m_2^2$ is ``natural.'' One thing is very clear from here
that if $\left| \mu _{12}^2\right| $ were much smallar, then neutrino masses
would be too small to account for the present observation of the neutrino
oscillations.

The $6\times 6$ mass matrix spaning $\left[ \nu _e\text{, }\nu _\mu \text{, }%
\nu _\tau \text{, }N_1\text{, }N_2\text{, }N_3\right] $ is now given by 
\begin{equation}
{\cal M}_\nu =\left[ 
\begin{array}{cccccc}
0 & 0 & 0 & f_{e1}u & f_{e2}u & f_{e3}u \\ 
0 & 0 & 0 & f_{\mu 1}u & f_{\mu 2}u & f_{\mu 3}u \\ 
0 & 0 & 0 & f_{\tau 1}u & f_{\tau 2}u & f_{\tau 3}u \\ 
f_{e1}u & f_{\mu 1}u & f_{\tau 1}u & M_1 & 0 & 0 \\ 
f_{e2}u & f_{\mu 2}u & f_{\tau 2}u & 0 & M_2 & 0 \\ 
f_{e3}u & f_{\mu 3}u & f_{\tau 3}u & 0 & 0 & M_3
\end{array}
\right] .  \label{4115}
\end{equation}
The mixing between $\nu $ and $N$ is thus of the order $fu/M$, which will
allow the physical $N$ to decay through its small component of $\nu $ to $%
l^{\pm }W^{\mp }$. The effective mass matrix spaning the light neutrino is
then 
\begin{equation}
{\cal M}_{ij}=\sum_k\frac{f_{ik}f_{jk}u^2}{M_k}.  \label{4116}
\end{equation}
and is of order $1$ $eV$ if $f$ is of order unity.

There are five physical Higgs boson with masses given by 
\begin{eqnarray}
m_{h^{\pm }}^2 &=&m_2^2+\lambda _3v^2+\left( \lambda _2-\lambda _4\right)
u^2-\mu _{12}^2u/v,  \label{4117} \\
m_A^2 &=&m_2^2+\left( \lambda _3+\lambda _4\right) v^2+\lambda _2u^2-\mu
_{12}^2u/v,  \label{4118} \\
m_{h_1^0}^2 &=&2\lambda _1v^2+{\cal O}\left( u^2\right) ,  \label{4119} \\
m_{h_2^0}^2 &=&m_2^2+\left( \lambda _3+\lambda _4\right) v^2+{\cal O}\left(
u^2\right) .  \label{4120}
\end{eqnarray}
The $m_{h_1^0}^2$ behaves very much like the ordionary Higgs boson. The new
scalar particles of this model, i.e. $h^{\pm }$, $A$, and $h_2^0$ (all with
mass$\sim m_2$), as well as $N_{iR}$, are now accessible to direct
experimental discovery in future accelators.

In summary, a new seesaw model of neutrino mass is proposed, where a second
scalar doublet $(\eta ^{+},\eta ^{0})$ with lepton number $L=-1$ is added to
the minimal Standard Model togather with three neutral right-handed fermion
singlets $N_{i}$ with lepton number $L=0$. Thus $N_{i}$ is allowed to have a
Majorana mass $m_{N}$ as well as interaction $f_{ij}\bar{N}_{iR}\left( \nu
_{jL}\eta ^{0}-l_{jL}\eta ^{+}\right) $. Hence $m_{\nu }$ is proportional to 
$\left\langle \eta ^{0}\right\rangle ^{2}/m_{i}$ and, if $\left\langle \eta
^{0}\right\rangle \ll \left\langle \phi ^{0}\right\rangle ,$ $m_{N}$ may be
of the order $1$ $TeV$ and be observable experimently. This is accomplished
with the Higgs potential of Eq. (\ref{4110}), where $L$ is broken explicitly
and uniquely with the soft term $\Phi ^{\dagger }\eta +\eta ^{\dagger }\Phi $%
.

As the Lepton Flavor Violation is discussed in all models of neutrino mass.
It is argued that in this model, there is no LFV at tree level for charged
leptons. However, it does occur in one loop throug $\eta $ and $N$ exchange 
\cite{PRL. 86 2502 (2001)}.

The aim of this work is to test this model for the lepton flavor violating
decays, i.e. $\mu \rightarrow 3e$ and $\mu ^{+}\rightarrow e^{+}\nu _l\bar{%
\nu}_l$ in one loop process. We have to calculate the branching ratio of $%
\mu \rightarrow 3e$ and compare it to its experimenal bounds.

\subsection{Basic consideration and calculation}

First of all we will discuss the reaction $\mu \rightarrow 3e$. The
corresponding box digrams can be shown in fig. \ref{figure}. The couplings
at each vertex can be taken from Eq. (\ref{4115}). Instead of $\gamma $
materices we have unity at each vertex, because each Higgs boson is a scalar.

Using Feynman rules, the amplitude can be written as follows \cite{Cheng
&Lee paper}: 
\begin{eqnarray}
iT\left( \mu \rightarrow 3e\right) &=&2\sum_i\left( f_{\mu
i}^{*}f_{ei}^{*}f_{ei}^{*}f_{ei}\right) \int \frac{d^4k}{\left( 2\pi \right)
^4}\bar{v}(p_1)\left[ \frac i{\not{k}-m_i}\right] v(p_2)  \nonumber \\
&&\times \bar{u}(p_3)\left[ \frac i{\not{k}-m_i}\right] v(p_4)\left[ \frac{-i%
}{k^2-m_h^2}\right] ^2  \nonumber \\
&=&2\sum_i\left( f_{\mu i}^{*}f_{ei}^{*}f_{ei}^{*}f_{ei}\right) \int \frac{%
d^4k}{\left( 2\pi \right) ^4}\bar{v}(p_1)\left[ \not{k}+m_i\right] v(p_2) 
\nonumber  \label{21} \\
&&\times \bar{u}(p_3)\left[ \not{k}+m_i\right] v(p_4)\left[ \frac
1{k^2-m_h^2}\right] ^2\left[ \frac 1{k^2-m_i^2}\right] ^2  \nonumber \\
&=&\frac 2{\left( 2\pi \right) ^4}\sum_i\left( f_{\mu
i}^{*}f_{ei}^{*}f_{ei}^{*}f_{ei}\right) \left[ \left\{ \bar{v}(p_1)\gamma
^\alpha v(p_2)\bar{u}(p_3)\gamma ^\beta v(p_4)I^{\alpha \beta }\right\}
\right.  \nonumber   \\
&&+m_i\left\{ \bar{v}(p_1)\gamma ^\alpha v(p_2)\bar{u}(p_3)v(p_4)I^\alpha +%
\bar{v}(p_1)v(p_2)\bar{u}(p_3)\gamma ^\beta v(p_4)I^\beta \right\}  \nonumber
\\
&&\left. +m_i^2\left\{ \bar{v}(p_1)v(p_2)\bar{u}(p_3)v(p_4)I\right\} \right]
,  \label{4121}
\end{eqnarray}
where 
\begin{eqnarray}
I^{\alpha \beta } &=&\int d^4k\frac{k^\alpha k^\beta }{\left(
k^2-m_h^2\right) ^2\left( k^2-m_i^2\right) ^2}  \label{4122} \\
I^\alpha &=&\int d^4k\frac{k^\alpha }{\left( k^2-m_h^2\right) ^2\left(
k^2-m_i^2\right) ^2}  \label{4123} \\
I &=&\int d^4k\frac 1{\left( k^2-m_h^2\right) ^2\left( k^2-m_i^2\right) ^2}
\label{4124}
\end{eqnarray}
It is assumed that the loop momenta is very high, i.e. $k\rightarrow \infty $%
. So there is no change in momenta at the vertices, and also the dominating
integral is the first one. Before calculating the $\left| T\right| ^2$, we
have to do the loop integration. To do this we use the technique known as
Feynman parameterization \cite{Mandal}.

\begin{equation}
\frac 1{a^2b^2}=6\int_0^1dz\frac{z\left( 1-z\right) }{\left[ b+(a-b)z\right]
^4}.  \label{4128}
\end{equation}

For our above mentioned case $a$ and $b$ are 
\begin{eqnarray}
a &=&k^2-m_i^2  \nonumber  \label{28} \\
b &=&k^2-m_h^2.  \label{4129}
\end{eqnarray}
So 
\begin{eqnarray}
\left[ b+(a-b)z\right] &=&k^2-m_h^2+\left( m_h^2-m_i^2\right) z  \nonumber \\
&=&k^2+m_h^2\left[ \left( 1-x_i\right) z-1\right]  \nonumber \\
&=&k^2+s,  \label{4127}
\end{eqnarray}
where $x_i=\left( \frac{m_i^2}{m_h^2}\right) $ and $s=m_h^2\left[ \left(
1-x_i\right) z-1\right] $. Then, Eq. (\ref{4122}) becomes 
\begin{eqnarray}
I^{\alpha \beta } &=&6\int_0^1dz\times z\left( 1-z\right) \int d^4k\frac{%
k^\alpha k^\beta }{\left[ k^2+s\right] ^4}  \nonumber \\
&=&6\int_0^1dz\times z\left( 1-z\right) \times \left[ \frac{i\pi ^2\Gamma
\left( 1\right) \times g^{\alpha \beta }}{2\Gamma \left( 4\right) \times s}%
\right]  \nonumber \\
&=&6\int_0^1dz\times z\left( 1-z\right) \times \left[ \frac{i\pi ^2g^{\alpha
\beta }}{2\times 6m_h^2\left[ \left( 1-x_i\right) z-1\right] }\right] ,
\label{41128}
\end{eqnarray}
because 
\[
\int d^4k\frac{k^\alpha k^\beta }{\left[ k^2+s+i\epsilon \right] ^n}=\frac{%
i\pi ^2\Gamma \left( n-3\right) }{2\Gamma \left( n\right) }\times \frac{%
g^{\alpha \beta }}{s^{n-3}}\qquad n\geq 3. 
\]
Therefore, after solving the $z$ integration Eq. (\ref{41128}) takes the
form 
\begin{equation}
I^{\alpha \beta }=\frac{i\pi ^2g^{\alpha \beta }}{2m_h^2}\times \left[ \frac{%
1-x_i^2+x_i\ln \left( x_i^2\right) }{2\left( x_i-1\right) ^3}\right] .
\label{41129}
\end{equation}
Using the same technique for Eq. (\ref{4123}) becomes 
\begin{eqnarray}
I^\alpha &=&\int d^4k\frac{k^\alpha }{\left( k^2-m_h^2\right) ^2\left(
k^2-m_i^2\right) ^2}  \nonumber \\
&=&\int d^4k\frac{k^\alpha }{\left( k^2-s\right) ^4}=0,  \label{4130}
\end{eqnarray}
because 
\[
\int d^4k\frac{k^\alpha }{\left[ k^2+s+i\epsilon \right] ^n}=0\qquad n\geq
3. 
\]
Also define 
\begin{equation}
f_{\mu i}^{*}f_{ei}^{*}f_{ei}^{*}f_{ei}=\xi _i.  \label{4131}
\end{equation}
Hence finally Eq. (\ref{4121}) becomes 
\begin{eqnarray}
T\left( \mu \rightarrow 3e\right) &=&\left. \frac{\pi ^2}{\left( 2\pi
\right) ^4m_h^2}\times \sum_i\xi _i\right[ \left\{ \bar{v}(p_1)\gamma
^\alpha v(p_2)\bar{u}(p_3)\gamma _\alpha v(p_4)\right\}  \nonumber
 \\
&&\left. \times \left\{ \frac{1-x_i^2+x_i\ln \left( x_i^2\right) }{2\left(
x_i-1\right) ^3}\right\} \right] .  \label{4132}
\end{eqnarray}
Now the square of the amplitude becomes 
\begin{eqnarray*}
\sum_{spins}\left| T\right| ^2 &=&\left( \frac{\pi ^2}{\left( 2\pi \right)
^4m_h^2}\right) ^2\sum_{i,j}\xi _i\xi _j^{*}\left\{ \frac{1-x_i^2+x_i\ln
\left( x_i^2\right) }{2\left( x_i-1\right) ^3}\right\} \\
&&\times \left\{ \frac{1-x_j^2+x_i\ln \left( x_j^2\right) }{2\left(
x_j-1\right) ^3}\right\} \\
&&\times 2\left[ \left( p_1p_4\right) \left( p_2p_3\right) +\left(
p_1p_3\right) \left( p_2p_4\right) \right] .
\end{eqnarray*}
Define 
\[
A\left( x_i\right) =\left\{ \frac{1-x_i^2+x_i\ln \left( x_i^2\right) }{%
2\left( x_i-1\right) ^3}\right\} . 
\]
Then the above result takes the form 
\begin{equation}
\sum_{spins}\left| T\right| ^2=\left[ \frac{\pi ^2}{\left( 2\pi \right)
^4m_h^2}\right] ^22\sum_{i,j}\xi _i\xi _j^{*}A\left( x_i\right) A\left(
x_j\right) \times \left[ \left( p_1p_4\right) \left( p_2p_3\right) +\left(
p_1p_3\right) \left( p_2p_4\right) \right] .  \label{4133}
\end{equation}
The corresponding decay width becomes 
\begin{equation}
\Gamma \left[ \mu \rightarrow 3e\right] =\left[ \frac{\pi ^2}{\left( 2\pi
\right) ^4m_h^2}\right] ^2\sum_{i,j}\xi _i\xi _j^{*}A\left( x_i\right)
A\left( x_j\right) \times \frac{m_\mu ^5}{192\times \pi ^3},  \label{4134}
\end{equation}
and the branching ratio becomes 
\begin{eqnarray}
{\cal B}\left[ \mu \rightarrow 3e\right] &=&\frac{\Gamma \left[ \mu
\rightarrow 3e\right] }{\Gamma \left[ \mu \rightarrow e\nu \bar{\nu}\right] }
\nonumber \\
&=&\left[ \frac 1{\left( 2\pi \right) ^42m_h^4}\right] \sum_{i,j}\xi _i\xi
_j^{*}A\left( x_i\right) A\left( x_j\right) \frac{M_W^6}{\Gamma ^2\left[
W\rightarrow e\nu \right] \times 2^5}.  \nonumber \\
&&  \label{4135}
\end{eqnarray}
Now let's calculate the other reaction, i.e. $\mu ^{+}\rightarrow e^{+}\nu _l%
\bar{\nu}_l.$

The corresponding box diagrams can be shown in fig. \ref{fig31}.

Using Feynman rules, the amplitude can be written as follows 
\begin{eqnarray}
iT\left( \mu ^{+}\rightarrow e^{+}\nu _l\bar{\nu}_l\right) &=&2\sum_i\left(
f_{\mu i}^{*}f_{li}^{*}f_{ei}^{*}f_{li}\right) \int \frac{d^4k}{\left( 2\pi
\right) ^4}\bar{v}(p_1)\left[ \frac i{\not{k}-m_i}\right]  \nonumber \\
&&\times v(p_2)\bar{u}(p_3)\left[ \frac i{\not{k}-m_i}\right] v(p_4)\left[ 
\frac{-i}{k^2-m_{h^{+}}^2}\right] \left[ \frac{-i}{k^2-m_{h_2^0}^2}\right] 
\nonumber \\
&=&2\sum_i\left( f_{\mu i}^{*}f_{li}^{*}f_{ei}^{*}f_{li}\right) \int \frac{%
d^4k}{\left( 2\pi \right) ^4}\bar{v}(p_1)\left[ \not{k}+m_i\right] v(p_2) 
\nonumber \\
&&\times \bar{u}(p_3)\left[ \not{k}+m_i\right] v(p_4)\frac
1{k^2-m_{h^{+}}^2}\left( \frac 1{k^2-m_i^2}\right) ^2\frac{-i}{%
k^2-m_{h_2^0}^2}.  \nonumber \\
&&  \label{4136}
\end{eqnarray}
Using the same assumption here that the loop moment is very high, i.e. $%
k\rightarrow \infty $. The above integral can be solvd by using Feynman
parameterization \cite{Cheng & Lie} 
\[
\frac 1{a^2bc}=6\int_0^1dx\int_0^xdy\frac{\left( 1-x\right) }{\left[
a+\left( b-a\right) x+\left( c-b\right) y\right] ^4} 
\]
Therefore 
\begin{eqnarray}
I^{\alpha \beta } &=&\int d^4k\frac{k^\alpha k^\beta }{\left(
k^2-m_i^2\right) ^2\left( k^2-m_{h^{+}}^2\right) \left(
k^2-m_{h_2^0}^2\right) }  \nonumber \\
&=&\frac{-i\pi ^2}{4m_i^2}A(x_{1i},x_{2i})g^{\alpha \beta },  \label{4137}
\end{eqnarray}
where 
\[
x_{1i}=\frac{m_{h^{+}}^2}{m_i^2},\qquad x_{2i}=\frac{m_{h_2^0,A}^2}{m_i^2}, 
\]
and 
\begin{eqnarray}
A(x_{1i},x_{2i}) &=&\frac{J\left( x_{1i}\right) -J\left( x_{2i}\right) }{%
x_{1i}-x_{2i}}  \nonumber \\
J\left( x_{1i}\right) &=&\frac 1{\left( 1-x_{1i}\right) }+\frac{x_{1i}^2\ln
x_{1i}}{\left( 1-x_{1i}\right) ^2}  \label{4138}
\end{eqnarray}
Define 
\begin{equation}
f_{\mu i}^{*}f_{li}^{*}f_{ei}^{*}f_{li}=\xi _i.  \label{4139}
\end{equation}
Using these results Eq. (\ref{4136}) takes the form 
\begin{eqnarray}
iT\left( \mu ^{+}\rightarrow e^{+}\nu _l\bar{\nu}_l\right) &=&\left[ \frac{%
-i\pi ^2}{2m_i^2\left( 2\pi \right) ^4}\right] \sum_i\xi _iA(x_{1i},x_{2i}) 
\nonumber   \\
&&\times \left\{ \bar{v}(p_1)\gamma ^\alpha v(p_2)\bar{u}(p_3)\gamma _\alpha
v(p_4)\right\} .  \label{4140}
\end{eqnarray}
Now, after calculating $\sum_{spins}\left| T\right| ^2$ , the decay width
becomes 
\begin{eqnarray}
\Gamma \left[ \mu ^{+}\rightarrow e^{+}\nu _l\bar{\nu}_l\right] &=&\left[ 
\frac{\pi ^2}{2m_i^2\left( 2\pi \right) ^4}\right] ^2\sum_{i,j}\xi _i\xi
_jA(x_{1i},x_{2i})A(x_{1j},x_{2j})\frac{m_\mu ^5}{192\times \pi ^3} 
\nonumber \\
&=&\left[ \frac 1{32m_i^2\pi ^2}\right] ^2\sum_{i,j}\xi _i\xi
_jA(x_{1i},x_{2i})A(x_{1j},x_{2j})\frac{m_\mu ^5}{192\times \pi ^3}. 
\nonumber \\
&&  \label{4141}
\end{eqnarray}
Then the branching ratio becomes 
\begin{equation}
{\cal B}\left[ \mu ^{+}\rightarrow e^{+}\nu _l\bar{\nu}_l\right] =\left[
\frac 1{32m_i^2\pi ^2}\right] ^2\sum_{i,j}\left( \xi _i\xi
_jA(x_{1i},x_{2i})A(x_{1j},x_{2j})\right) \frac{M_W^6}{2\times \Gamma
^2\left[ W\rightarrow e\nu \right] }.  \label{4142}
\end{equation}
In order to calculate the numerical value of the branching ratio for $\mu
\rightarrow 3e$, we will fix the value of $m_i=1$ TeV and also assume that
all the couplings are of the order unity. As $m_h^2\sim m_2^2$, therefore if 
$m_2\gg 100$ GeV$\,$ its effect on the radiative parameters 
\begin{eqnarray*}
\Delta S &=&\frac 1{24\pi }\frac{\lambda _4v^2}{m_2^2} \\
\Delta T &=&\frac 1{96\pi }\frac 1{s^2c^2M_Z^2}\frac{\lambda _4^2v^4}{m_2^2},
\end{eqnarray*}
is negligible small and will not change the excellent experimental fit of
the minimal Standard Model.

To have a feeling about the branching ratios let us take $m_h=900$ GeV, the
branching ratio for $\mu \rightarrow 3e$ becomes 
\[
{\cal B}=3.6\times 10^{-6} 
\]
which is much higher then the experimental bound on the branching ratio of $%
\mu \rightarrow 3e$, i.e. ${\cal B}\left( \mu \rightarrow 3e\right)
_{V-exch.}\leq 10^{-12}$. For the mass smallar than $900$ GeV the branching
ratio grows higher and higher. Thus from here it is obvious that $m_h>1$
TeV. By varing the Higgs mass the corresponding branching ratio becomes

\[
\begin{array}{cc}
m_h=10TeV & {\cal B}=1.0\times 10^{-8} \\ 
m_h=30TeV & {\cal B}=3.1\times 10^{-11}.
\end{array}
\]
It can be eaisly seen that with the increase of Higgs mass, the branching
ratio decreases. As the Higgs mass reaches to $50$ TeV, branching ratio
becomes 
\[
{\cal B}=3.6\times 10^{-12} 
\]
which is comparable to the experimental value. Thus we have concluded that
the Higgs mass is greater than $50$ TeV to explain LFV decays in Seesaw
model of neutrino masses \cite{PRL. 86 2502 (2001)}.

\chapter*{Conclusion}

The aim of this work has been to understand the physics beyond the Standard
Model by considering Lepton Flavour Violating (LFV) decay processes. Using
the experimental bounds on the three body LFV decays, we have calculated the
bounds on the corresponding two body decays. The Dynamical suppression of
three boby LFV decays due to momentum dependent couplings have also been
discussed in detail.

The experimental bounds on the three body LFV decay, $\mu \rightarrow 3e$
constrained the coupling $\tilde{g}_{Z_{\mu e}}$, which is helpful to
calculate the anamolous muon decays. The anamolous muon decay $\mu
\rightarrow e\nu _l\bar{\nu}_l$ could not significantly contribute to the
LSND DAR result. This has been concluded by comparing the effective coupling
of anamolous muon decay with $r>1.6\times 10^{-3}$ \cite{hep-ph/9809524}.
LFV decays at a loop order have also been discussed in Seesaw model of
neutrino masses which involve right handed singlet neutrinos of mass $m_N$
of order TeV \cite{PRL. 86 2502 (2001)}. The main purpose of the model is
that there is no new physics beyond $1$ TeV. It has been shown that even if
one keeps $m_N$ at $1$ TeV the present bounds on $\mu \rightarrow 3e$
requires that new Higgs bosons, necessary in this model should have their
mass larger than $50$ TeV. Thus it is necessary to have a new physics beyond
the TeV energy scale defeating the origional motivation of the model.

\newpage 

%
%
%
%

\begin{figure}
\label{tree4}
\caption{Anamolous muon decay}
\end{figure}

\begin{figure}
\label{figure}
\caption{Box diagram for $\mu \rightarrow 3 e$}
\end{figure}

\begin{figure}
\label{fig31}
\caption{Box diagram for anamolous muon decay}
\end{figure}
%

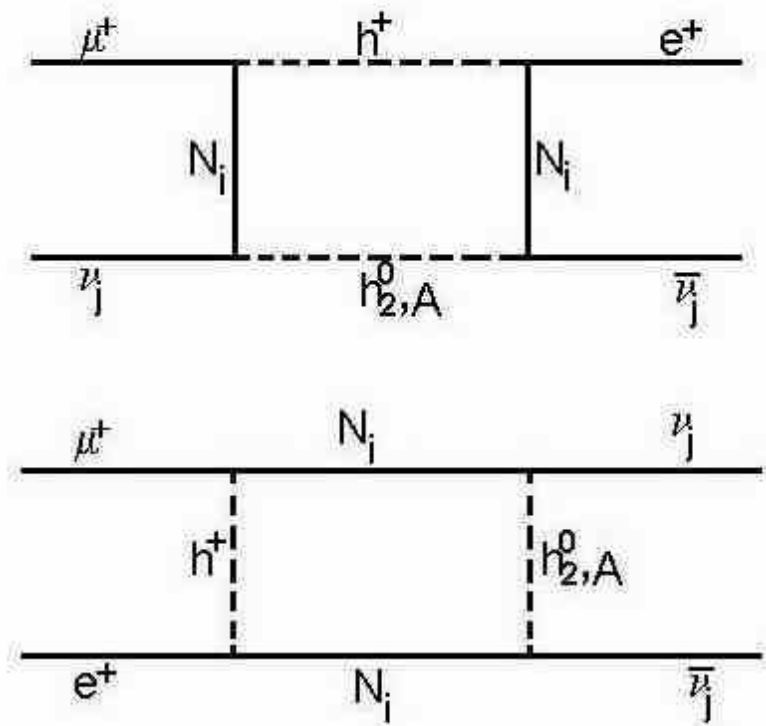
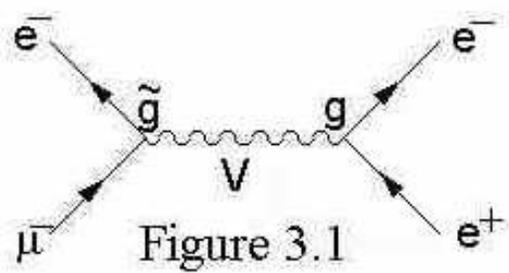
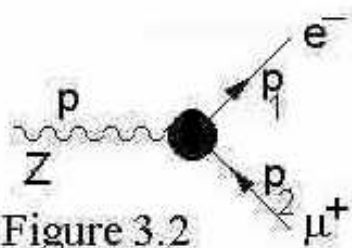
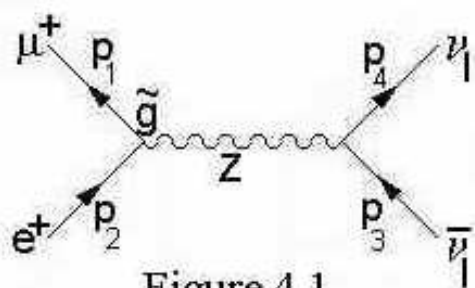
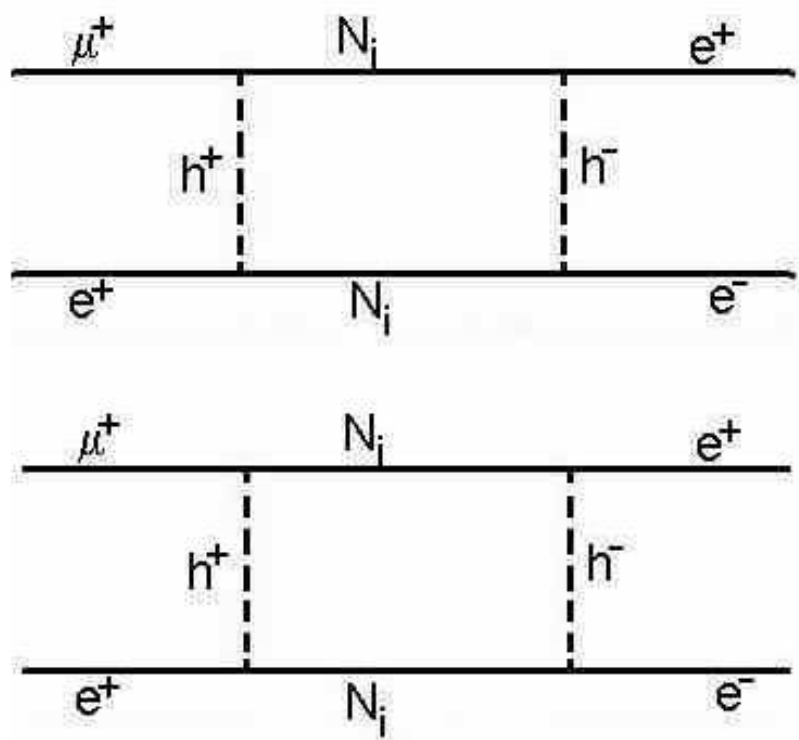
\end{document}